\documentclass[apj]{emulateapj}
\usepackage{graphicx}
\usepackage{amsmath}
\usepackage{hyperref}

\newcommand{\hi}{\mbox{\rm H$\,$\scshape{i}}}
\newcommand{\hii}{\mbox{\rm H$\,$\scshape{ii}}}
\newcommand{\nii}{\mbox{\rm [N$\,$\scshape{ii}]}}
\newcommand{\htwo}{\mbox{\rm H$_2$}}
\newcommand{\ha}{\mbox{\rm H$\alpha$}}
\newcommand{\jone}{\mbox{($1\rightarrow0$)}}
\newcommand{\jtwo}{\mbox{($2\rightarrow1$)}}
\newcommand{\kmpers}{\mbox{km~s$^{-1}$}}
\newcommand{\Kkmpers}{\mbox{K~km~s$^{-1}$}}
\newcommand{\xco}{\mbox{$X_{\rm CO}$}}
\newcommand{\xcounits}{\mbox{cm$^{-2}$ (K km s$^{-1}$)$^{-1}$}}
\newcommand{\MJypersr}{\mbox{MJy~sr$^{-1}$}}
\newcommand{\Msunperpc}{\mbox{\rm M$_{\odot}$ pc$^{-2}$}}
\newcommand{\Msunperyrperkpc}{\mbox{\rm M$_{\odot}$ yr$^{-1}$ kpc$^{-2}$}}

\shorttitle{SF Law in HI--Dominated Regime}
\shortauthors{Schruba et al.}
\slugcomment{Accepted for publication in \emph{The Astronomical Journal}}

\begin{document}
\title{A Molecular Star Formation Law in the Atomic Gas Dominated Regime in Nearby Galaxies}

\author{
Andreas Schruba\altaffilmark{1},
Adam K. Leroy\altaffilmark{2,11},
Fabian Walter\altaffilmark{1},
Frank Bigiel\altaffilmark{3},
Elias Brinks\altaffilmark{4},
W.\,J.\,G. de Blok\altaffilmark{5},\newline % added to get a nicer line break
Gaelle Dumas\altaffilmark{1},
Carsten Kramer\altaffilmark{6},
Erik Rosolowsky\altaffilmark{7},
Karin Sandstrom\altaffilmark{1},
Karl Schuster\altaffilmark{8},
Antonio Usero\altaffilmark{9},
Axel Weiss\altaffilmark{10},
Helmut Wiesemeyer\altaffilmark{10}}
\altaffiltext{1}{Max-Planck-Institut f\"ur Astronomie, K\"onigstuhl 17, 69117 Heidelberg, Germany; schruba@mpia.de}
\altaffiltext{2}{National Radio Astronomy Observatory, 520 Edgemont Road, Charlottesville, VA 22903, USA}
\altaffiltext{3}{Department of Astronomy, Radio Astronomy Laboratory, University of California, Berkeley, CA 94720, USA}
\altaffiltext{4}{Centre for Astrophysics Research, University of Hertfordshire, Hatfield AL10 9AB, U.K.}
\altaffiltext{5}{Astrophysics, Cosmology and Gravity Centre, Department of Astronomy, University of Cape Town, Private Bag X3, Rondebosch 7701, South Africa}
\altaffiltext{6}{IRAM, Avenida Divina Pastora 7, 18012 Granada, Spain}
\altaffiltext{7}{Department of Physics and Astronomy, University of British Columbia Okanagan, 3333 University Way, Kelowna, BC V1V 1V7, Canada}
\altaffiltext{8}{IRAM, 300 rue de la Piscine, 38406 St. Martin d\textquoteright H\`{e}res, France}
\altaffiltext{9}{Observatorio Astron\'{o}mico Nacional, C/ Alfonso XII, 3, 28014, Madrid, Spain}
\altaffiltext{10}{MPIfR, Auf dem H\"ugel 69, 53121 Bonn, Germany}
\altaffiltext{11}{Hubble Fellow}

\begin{abstract}
We use the IRAM HERACLES survey to study CO emission from 33
nearby spiral galaxies down to very low intensities. Using 21-cm line
atomic hydrogen (\hi) data, mostly from THINGS, we predict the local
mean CO velocity based on the mean \hi\ velocity. By re--normalizing
the CO velocity axis so that zero corresponds to the local mean \hi\
velocity we are able to stack spectra coherently over large regions.
This enables us to measure CO intensities with high significance as
low as $I_{\rm CO} \approx 0.3$ \Kkmpers\ ($\Sigma_{\rm H2} \approx
1$~\Msunperpc), an improvement of about one order of magnitude over
previous studies. We detect CO out to galactocentric radii $r_{\rm gal}
\sim r_{\rm 25}$ and find the CO radial profile to follow a remarkably
uniform exponential decline with scale length of $\sim$$0.2~r_{25}$.
Here we focus on stacking as a function of radius, comparing our
sensitive CO profiles to matched profiles of \hi , \ha , FUV, and IR
emission at 24~\micron\ and 70~\micron. We observe a tight, roughly
linear relationship between CO and IR intensity that does not show
any notable break between regions that are dominated by molecular
gas ($\Sigma_{\rm H2} > \Sigma_{\rm HI}$) and those dominated by
atomic gas ($\Sigma_{\rm H2} < \Sigma_{\rm HI}$). We use combinations
of FUV$+$24\micron\ and \ha$+$24\micron\ to estimate the recent star
formation rate (SFR) surface density, $\Sigma_{\rm SFR}$, and find
approximately linear relations between $\Sigma_{\rm SFR}$ and
$\Sigma_{\rm H2}$. We interpret this as evidence for stars forming
in molecular gas with little dependence on the local total gas surface
density. While galaxies display small internal variations in the
SFR--to--H$_2$ ratio, we do observe systematic galaxy--to--galaxy
variations. These galaxy--to--galaxy variations dominate the scatter in
relationships between CO and SFR tracers measured at large scales.
The variations have the sense that less massive galaxies exhibit larger
ratios of SFR--to--CO than massive galaxies. Unlike the SFR--to--CO
ratio, the balance between atomic and molecular gas depends strongly
on the total gas surface density and galactocentric radius. It must also
depend on additional parameters. Our results reinforce and extend to
lower surface densities a picture in which star formation in galaxies is
separable into two processes: the assembly of star--forming molecular
clouds and the formation of stars from H$_2$. The interplay between
these processes yields a total gas--SFR relation with a changing slope,
which has previously been observed and identified as a star formation
threshold.
\end{abstract}


\section{Introduction}
\label{sec:intro}

Stars form out of molecular (\htwo) gas and many recent observations
of nearby galaxies have revealed a strong correlation between the
surface density of molecular gas, $\Sigma_{\rm H2}$, and the star
formation rate (SFR) surface density, $\Sigma_{\rm SFR}$
(\citealt{Wong2002, Kennicutt2007, Bigiel2008, Leroy2008,
Wilson2008, Blanc2009}; see also the recent review by
\citealt{Bigiel2011}). These studies show a correlation over several
orders of magnitude, but mostly for regions where \htwo\ makes up the
majority of the neutral gas, $\Sigma_{\rm H2} \gtrsim \Sigma_{\rm HI}$.
The lack of a clear correlation between atomic gas (\hi) surface density,
$\Sigma_{\rm HI}$, and $\Sigma_{\rm SFR}$ inside galaxy disks
\citep[e.g.,][]{Bigiel2008} offers circumstantial evidence that star
formation remains coupled to the molecular, rather than total
(\hi+\htwo), gas surface density even where \hi\ makes up most of the
interstellar medium (ISM). However, the exact relationship between
$\Sigma_{\rm SFR}$ and $\Sigma_{\rm H2}$ in the \hi --dominated parts
of galaxies ($\Sigma_{\rm H2} \lesssim \Sigma_{\rm HI}$) remains
largely unexplored.

In this paper we use new, sensitive, wide--field CO maps from the
IRAM\footnote{IRAM is supported by CNRS/INSU (France), the MPG
  (Germany) and the IGN (Spain).} HERACLES survey \citep{Leroy2009a}
to measure correlations between molecular gas and SFR tracers over a
large dynamic range.  By employing stacking techniques based on
\hi\ priors we extend our observations from H$_2$--dominated galaxy
centers to the outer parts of galaxies where the H$_2$ surface density
is much lower than the \hi\ surface density, $\Sigma_{\rm H2} \ll
\Sigma_{\rm HI}$.

Deep CO measurements allow us to test if a single ``star formation
law'' applies in both the \htwo -- and \hi --dominated parts of
galaxies.  Following \citet{Schmidt1959}, astronomers have
investigated scaling relations linking gas and star formation for
decades. Such relations only approximate the complex physical
processes involved in star formation but provide useful constraints
on theoretical models and important input to simulations. After
\citet{Kennicutt1989, Kennicutt1998}, power laws linking surface
densities of gas and SFR are the most common formulation. However,
the choice of which gas surface density to use --- total or molecular gas
--- remains controversial, as does the extension of any measured
molecular relation to low surface densities. The underlying question
is what limits star formation in low column density regions, the
formation of molecular gas or the efficiency at which the available
molecular gas is converted into stars? Sensitive observations of
molecular gas down to low surface densities in a large sample of
galaxies are needed to address these questions.

Our CO measurements also allow us to investigate the distribution of
molecular gas out to large radii. A characteristic exponential decline
has been observed several times \citep{Young1995, Regan2001,
  Leroy2009a}, but it is not known if this decline becomes sharper
at one point, for example corresponding to claimed star formation
thresholds \citep[e.g.,][]{Martin2001}. We also test how variations
in the \htwo--to--\hi\ ratio extend to low surface densities. This
quantity is a strong and systematic function of environment in nearby
galaxies \citep[][]{Wong2002, Blitz2006, Leroy2008, Hitschfeld2009}
but it has been difficult to extend the observed correlations to low
surface densities.

In Section~\ref{sec:sample_data} we describe our sample and data.
In Section~\ref{sec:method} we present the method that we use to extract
sensitive CO measurements. In Section~\ref{sec:results} we present
radial profiles of \hi, CO, IR, FUV, and \ha\ and use these to relate CO, 
\hi , and tracers of recent star formation. In Section~\ref{sec:summary}
we summarize our results.

\section{Sample \& Data}
\label{sec:sample_data}

We study 33 nearby, star--forming disk galaxies, the set of HERACLES
targets for which we could collect the necessary \hi, IR, FUV, and \ha\
data. This is mainly the overlap of several surveys: HERACLES \citep[IRAM
  30m CO,][]{Leroy2009a}, THINGS \citep[VLA \hi,][] {Walter2008}, SINGS
or LVL \citep[{\em Spitzer} IR,][]{Kennicutt2003b, Dale2009}, and the
GALEX NGS \citep[GALEX FUV,][]{GildePaz2007}. We supplement
these with a combination of archival and new \hi\ data and archival
GALEX data. We exclude low mass, low metallicity galaxies with only
upper limits on CO emission and nearly edge--on galaxies.

Table~\ref{t1} lists our sample along with adopted morphology, distance,
inclination, position angle, optical radius, and metallicity. These values
are taken from \citet{Walter2008} if possible and from LEDA and NED
in other cases. We quote oxygen abundances (metallicities) from
\citet[][Table 9]{Moustakas2010}, averaging the metallicities derived from
a theoretical calibration (their KK04 values) and an empirical calibration
(their PT05 values). For galaxies without a \citet{Moustakas2010}
metallicity, we adopt a metallicity equal to the average of their
\mbox{B--band} luminosity--metallicity relations. For NGC~2146 we
quote the metallicity given by \citet{Engelbracht2008}. For NGC~5457
(M~101) we take a constant metallicity defined by the value at $0.4~r_{25}$
from the gradient fit by \citet{Kennicutt2003a}.

\begin{deluxetable}{llrrrrc}
\tablecolumns{7}
\tablecaption{Properties of Galaxy Sample\label{t1}}
\tablehead{\colhead{Galaxy} & \colhead{Morph.} & \colhead{$D$} & \colhead{Incl.} & \colhead{P.A.} & \colhead{$r_{\rm 25}$} & \colhead{Metal.\tablenotemark{c}} \\
\colhead{} & \colhead{} & \colhead{Mpc} & \colhead{$^{\circ}$} & \colhead{$^{\circ}$} & \colhead{$\arcmin$} & \colhead{12+logO/H}}
\startdata
NGC337 & SBd & 24.7 &  51 &  90 &  1.48 & 8.51 \\
NGC628\tablenotemark{a,b} & Sc &  7.3 &   7 &  20 &  4.92 & 8.68 \\
NGC925\tablenotemark{a,b} & SBcd &  9.2 &  66 & 287 &  5.32 & 8.52 \\
NGC2146 & SBab & 12.8 &  54 & 123 &  2.69 & 8.68\tablenotemark{d} \\
NGC2403\tablenotemark{a} & SBc &  3.2 &  63 & 124 &  7.87 & 8.57 \\
NGC2798\tablenotemark{a} & SBa & 24.7 &  85 & 152 &  1.20 & 8.69 \\
NGC2841\tablenotemark{a,b} & Sb & 14.1 &  74 & 153 &  3.45 & 8.87 \\
NGC2903\tablenotemark{a,b} & SBd &  8.9 &  65 & 204 &  5.92 & 8.90 \\
NGC2976\tablenotemark{a,b} & Sc &  3.6 &  65 & 335 &  3.60 & 8.67 \\
NGC3049 & SBab &  8.9 &  58 &  28 &  1.04 & 8.82 \\
NGC3077\tablenotemark{a} & Sd &  3.8 &  46 &  45 &  2.70 & 8.64 \\
NGC3184\tablenotemark{a,b} & SBc & 11.1 &  16 & 179 &  3.70 & 8.83 \\
NGC3198\tablenotemark{a,b} & SBc & 13.8 &  72 & 215 &  3.24 & 8.62 \\
NGC3351\tablenotemark{a,b} & SBb & 10.1 &  41 & 192 &  3.60 & 8.90 \\
NGC3521\tablenotemark{a,b} & SBbc & 10.7 &  73 & 340 &  4.16 & 8.70 \\
NGC3627\tablenotemark{a} & SBb &  9.3 &  62 & 173 &  5.14 & 8.67 \\
NGC3938 & Sc & 12.2 &  14 &  15 &  1.77 & 8.74 \\
NGC4214\tablenotemark{a,b} & Irr &  2.9 &  44 &  65 &  3.40 & 8.25 \\
NGC4254 & Sc & 20.0 &  32 &  55 &  2.51 & 8.79 \\
NGC4321 & SBbc & 14.3 &  30 & 153 &  3.01 & 8.83 \\
NGC4536 & SBbc & 14.5 &  59 & 299 &  3.54 & 8.61 \\
NGC4559 & SBcd & 11.6 &  65 & 328 &  5.24 & 8.55 \\
NGC4569 & SBab & 20.0 &  66 &  23 &  4.56 & 8.92 \\
NGC4579 & SBb & 20.6 &  39 & 100 &  2.51 & 8.88 \\
NGC4625 & SBmp &  9.5 &  47 & 330 &  0.69 & 8.70 \\
NGC4725 & SBab &  9.3 &  54 &  36 &  4.89 & 8.73 \\
NGC4736\tablenotemark{a,b} & Sab &  4.7 &  41 & 296 &  3.87 & 8.66 \\
NGC5055\tablenotemark{a,b} & Sbc & 10.1 &  59 & 102 &  5.93 & 8.77 \\
NGC5194\tablenotemark{a} & SBc &  8.0 &  20 & 172 &  3.85 & 8.86 \\
NGC5457\tablenotemark{a} & SBcd &  7.4 &  18 &  39 & 11.99 & 8.46\tablenotemark{e} \\
NGC5713 & Scd & 26.5 &  48 &  11 &  1.23 & 8.64 \\
NGC6946\tablenotemark{a,b} & SBc &  5.9 &  33 & 243 &  5.70 & 8.72 \\
NGC7331\tablenotemark{a,b} & Scd & 14.7 &  76 & 168 &  4.59 & 8.68
\enddata
\tablenotetext{a}{Targets of THINGS survey \citep{Walter2008}}
\tablenotetext{b}{Targets in first HERACLES survey paper \citep{Leroy2009a}}
\tablenotetext{c}{Oxygen abundance from \citet{Moustakas2010}}
\tablenotetext{d}{Oxygen abundence from \citet{Engelbracht2008}}
\tablenotetext{e}{Oxygen abundance from \citet{Kennicutt2003a}}
\end{deluxetable}

We trace molecular hydrogen (H$_2$) using CO$\jtwo$ line
emission observed with the IRAM 30m as part of the HERACLES
survey \citep{Leroy2009a}. They describe in detail the observations
and reduction for the subset of galaxies observed until Summer 2008.
The remaining targets were observed and reduced in the same way.
The final data cubes have an angular resolution (FWHM) of 13\arcsec\
and a spectral resolution (channel separation) of $2.6$~\kmpers.

Our measurements of atomic hydrogen (\hi ) come mostly from the
THINGS survey \citep{Walter2008}, which used the Very Large Array
\footnote{The National Radio Astronomy Observatory is a facility of the
National Science Foundation operated under cooperative agreement
by Associated Universities, Inc.}
(VLA) to observe the 21-cm hydrogen line in 34 nearby galaxies. The
observing and reduction strategies are described therein. The final
data cubes have an angular resolution of $\sim$11\arcsec\ (using
natural weighting) and a spectral resolution of $2.6$ or $5.2$ \kmpers.
THINGS is sensitive to $\Sigma_{\rm HI} \approx 0.5$ \Msunperpc\ 
on scales of 30\arcsec . Using azimuthal averaging, we reach even
better sensitivities at large radii. As a result, the \hi\ sensitivity never
limits our analysis.

The \hi\ data for NGC 337, 2146, 2798, 3049, 3938, 4254, 4321,
4536, 4579, 4625, 4725, and 5713 are a combination of new and
archival VLA data (the new data are from VLA programs AL731 and
AL735). These have been reduced and imaged using the Common
Astronomy Software Applications (CASA) following a similar protocol
than the THINGS reduction. These supplemental \hi\ cubes include
only data from the VLA's C and D configurations; THINGS also
includes B configuration data. For NGC 4559 we take \hi\ data
observed with the Westerbork Synthesis Radio Telescope (WSRT) by
\citet{vanderHulst2002}. The beam sizes (FWHM) of the supplemental
\hi\ are $15\arcsec - 25$\arcsec\ and the velocity resolution is $2.5 - 20$
\kmpers , usually $10$ \kmpers .

We derive \hi\ surface densities from 21-cm line intensities and
H$_2$ surface densities from CO\jtwo\ line intensities following

\begin{eqnarray}
	\Sigma_{\rm HI} &=& 0.02~I_{\rm HI} \times \cos~i \label{eq:hi}\\[2mm]
	\Sigma_{\rm H2} &=& 6.25~I_{\rm CO} \times \cos~i \label{eq:h2}
\end{eqnarray}

\noindent where $\Sigma_{\rm HI}$ and $\Sigma_{\rm H2}$ have
units of \Msunperpc\ and $I_{\rm HI}$ and $I_{\rm CO}$ are measured
in \Kkmpers. The mass surface densities are projected to face--on
values and include a factor of $1.36$ to account for heavy elements.
For Equation~\eqref{eq:h2} we have assumed a CO line ratio of 
$I_{\rm CO} \jtwo / I_{\rm CO} \jone = 0.7$ and a CO\jone--to--H$_2$
conversion factor $\xco = 2.0 \times 10^{20}$ \xcounits\ \citep[see][and
references therein]{Leroy2009a}.

We use broadband infrared (IR) photometry at 24~\micron\ and
70~\micron\ obtained by the {\em Spitzer} legacy surveys SINGS
\citep{Kennicutt2003a} and LVL \citep{Dale2009}. {\em Spitzer} has
angular resolution $\sim$6\arcsec\ at 24~\micron\ and 18\arcsec\ at
70~\micron . The sensitivity of these data is sufficient to obtain high
signal--to--noise measurements when averaging in radial rings
(see below) throughout the area that we study ($r_{\rm gal} \lesssim
1.2~r_{25}$).

The GALEX NGS \citep{GildePaz2007} imaged far-- and near--ultraviolet
(FUV and NUV) emission for most of our targets. The FUV band covers
$1350 - 1750$~\AA\ with an angular resolution $\sim$4.5\arcsec. We
use these images to trace unobscured emission from young stars. For
galaxies not covered by the NGS, we searched the NASA Multimission
Archive at STScI and used the FUV image with the longest exposure
time. These data also have sufficient sensitivity to determine FUV
intensities with high signal--to--noise throughout the star--forming disk.

We draw \ha\ data from the SINGS and LVL surveys, complemented
by literature data for NGC~2903, 4214, 4569, 4736, and 5457. For
the literature and several problematic SINGS targets, we pin the total
\ha$+$\nii\ flux to published values, usually those of \citet{Kennicutt2008}.
Leroy et al. (2011, in prep.) describe the processing of the maps, which
involves subtracting a smooth background, masking foreground stars
following \citet{MunozMateos2009}, correcting for \nii\ contamination
following \citet{Kennicutt2008}, and correcting for Galactic extinction.
The \ha\ maps become uncertain, and likely biased low due to
background subtraction, below intensities equivalent to $\Sigma_{\rm
SFR}$ of a few times $10^{-4}$~\Msunperyrperkpc , values typically
crossed inside the radial range studied here. This, and the declining
\ha--to--FUV flux ratios which are observed as {\em Galex} UV disks
extending far beyond the \ha\ emission \citep{Thilker2007, Meurer2009},
limit the utility of the \ha\ maps to trace star formation in the low brightness
regions of outer galaxy disks.

After we examine correlations among observables, we will estimate
the star formation rate surface density, $\Sigma_{\rm SFR}$, from
combinations of H$\alpha$ with 24~\micron\ \citep{Kennicutt2007}
and FUV with 24~\micron\ \citep{Leroy2008}. We adopt the
\ha$+$24\micron\ calibration by \citet{Calzetti2007} and the
FUV$+$24\micron\ combination from \citet{Leroy2008}, 

\begin{eqnarray}
	\Sigma_{\rm SFR(H\alpha+24)} &=& \nonumber\\
	&& \hspace{-18mm} \left( 2.9\! \times\! 10^{-2}\ I_{\rm H\alpha} + 2.5\! \times\! 10^{-3}\ I_{\rm 24 \mu m} \right) \times \cos~i \label{eq:sfr_ha24} \\[2mm]
	\Sigma_{\rm SFR(FUV+24)} &=& \nonumber\\
	&& \hspace{-2cm} \left( 8.1\! \times\! 10^{-2}\ I_{\rm FUV} + 3.2\! \times\! 10^{-3}\ I_{\rm 24\mu m} \right) \times \cos~i \label{eq:sfr_fuv24}
\end{eqnarray}

\noindent where $\Sigma_{\rm SFR}$ has units of \Msunperyrperkpc\
and \ha, FUV, and 24~\micron\ intensities are all in \MJypersr . Both
$\Sigma_{\rm SFR}$ calibrations combine a tracer of the unobscured
star formation with infrared (24~\micron) emission, which is intended
to trace young starlight reprocessed by dust. \ha\ traces O stars with
ages $\lesssim$5~Myrs \citep{Kennicutt2009} with sensitivity out to
$\sim$10~Myr \citep{Vacca1996}. FUV traces O and B stars of typical
age $20 - 30$ Myrs with sensitivity out to $\sim$100~Myr \citep{Salim2007}.

Our use of both \ha$+$24\micron\ and FUV$+$24\micron\ emission
gives some test of sensitivity to our choice of star formation rate tracer.
Several other concerns are worth mentioning. Because dust properties
and the stellar populations heating the dust somewhat differ between
\hii\ regions and large (kpc) regions in galaxies the appropriate
weighting of the 24~\micron\ emission to correct for extinction may
be a function of scale and environment. The dust--to--gas ratio, dust
size--distribution, ISM geometry, and recent star formation history
may also play important roles. The \citet{Calzetti2007} calculation
remains state of the art, but there is no definitive consensus about
the correct calibration to use outside bright regions that they study.
The reliability of the H$\alpha$ imaging at low surface brightness
also represents a concern. Ground based, narrowband \ha\ imaging
is challenging and the \ha$+$24\micron\ tracer must be considered
unreliable where $\Sigma_{\rm SFR(H\alpha)} \approx 5\times10^{-4}$
\Msunperyrperkpc . The {\em Galex} and {\em Spitzer} maps are
better behaved at low surface brightness.

Because of these data quality considerations and our focus on
regions with low surface brightness, we emphasize comparisons to
FUV$+$24\micron . SFR(FUV$+$24) and SFR(\ha$+$24) give
comparable results with scatter of only $0.1$~dex ($\sim$25\%)
down to  $\Sigma_{\rm SFR} \sim 5 \times 10^{-4}$  \Msunperyrperkpc\ 
in an azimuthally averaged ring. Our main method to address these
other systematic concerns is to emphasize the observed scaling
relations in the first part of the paper. For a more  thorough discussion
of hybrid SFR tracers see \citet{Kennicutt2009} and for a discussion
of their application to gas--SFR comparisons, we refer the reader
to Leroy et al. (2011, in prep.).

\section{Methodology}
\label{sec:method}

Our goal is to recover low brightness CO emission from the outer
parts of galaxies. CO is very faint in these regions and individual
spectra have low signal--to--noise ratios (SNR), requiring us to
average many spectra to achieve a detection. Because the velocity
of CO emission varies with position, simply averaging spectra spreads
the emission across many velocity channels with low SNR in each
channel. In principle, this could still yield a high SNR measurement.
In practice we wish to maximize SNR by considering only the part
of each spectrum likely to contain emission. We must also contend
with systematic effects of weather, receiver instabilities, and dish
imperfections. These all induce frequency--dependent behavior
(``baselines problems'') that make a clear detection of an emission
line an important step in a robust analysis.

\begin{figure}
\epsscale{1.0} \plotone{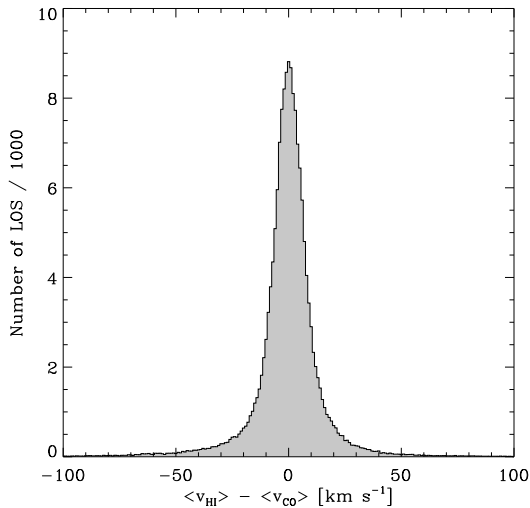}
\caption{The difference between local mean velocities of \hi\ and CO for
  lines of sight with galactocentric radius smaller than $0.5$~$r_{\rm 25}$.
  The close correspondence motivates our use of the \hi\ mean velocity
  to predict the CO velocities in low surface brightness regions.\label{f1}}
\end{figure}

With these issues in mind, we use the following technique to average
CO spectra across large parts of a galaxy. First, we estimate the
local mean velocity of CO emission from the \hi\ data. Using this mean
velocity, we redefine the velocity axis of each CO spectrum so that
the local mean velocity is now zero. We average these shifted CO
spectra from across our target region. In the averaged spectra we
expect the CO line to emerge at zero velocity with good SNR. The
baseline problems described above will not average coherently,
allowing a straightforward identification of the line.

This approach hinges on the assumption that the \hi\ mean velocity is
a good proxy for the mean velocity of the molecular gas. Figure~\ref{f1}
shows that this assumption holds where we detect CO over individual
lines of sight, with median $\bar{v}_{\rm CO} - \bar{v}_{\rm HI}$ of $-0.22$
\kmpers\ and a $1\sigma$ dispersion of $7.0$ \kmpers . We expect a
similar correspondence in the outer, CO--faint parts of galaxies.

Leveraging \hi\ to detect CO at large radii works because the \hi\ surface
densities are essentially constant out to large radii, making \hi\ easily
detected across galaxy disks. CO emission, on the other hand, tends
to be bright in galaxy centers but declines rapidly with increasing
galactocentric radius.

\subsection{Stacking of CO Spectra}
\label{sec:stacking}

Predicting the velocity of the CO line from the \hi\ data allows us to
increase the SNR when measuring the integrated CO intensity. The
HERACLES bandpass is $\sim$1000 \kmpers\ and a typical CO line
width at large galactocentric radii is $\sim$25 \kmpers.  The \hi\
allows us to restrict our integration to just the relevant part of the
spectrum which represents a substantial gain in sensitivity.

Just as important as the increase in SNR, the shifted and stacked
spectra allow us to verify that faint emission is actually an astronomical
signal. Even for faint CO emission the stacking technique has the
potential to reveal a spectral line. Low--level variations due to weather,
receiver instabilities, and other systematic effects in the telescope will
not create such an effect. Even stray pickup of astronomical emission
due to surface imperfections (i.e., error beam effects) will emerge at
a low level offset from zero velocity due to galaxy rotation.

\begin{figure}
\epsscale{1.1} \plotone{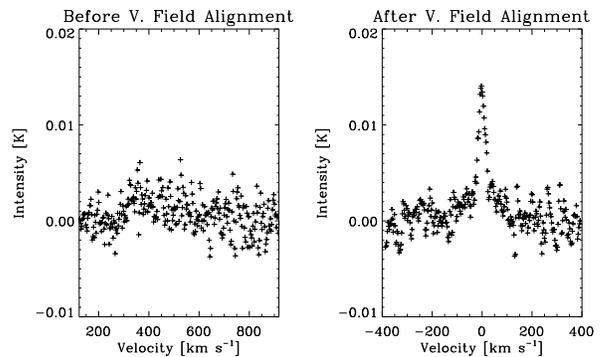}
\caption{Average CO spectrum across tilted ring spanning from
  $0.7 - 0.8$ $r_{\rm 25}$ in NGC 5055. The left panel shows the
  result of a simple average of all spectra. The right panel shows
  the average after each spectrum is shifted so that $v=0$
  corresponds to the local mean \hi\ velocity.\label{f2}}
\end{figure}

Figure~\ref{f2} demonstrates this approach. We plot the averaged
CO spectrum of NGC 5055 inside a tilted ring spanning from $0.7 - 0.8$
$r_{\rm 25}$. In the left panel the spectra were averaged as they were
observed, whereas in the right panel we first shifted by the local \hi\ mean
velocity and then averaged. Both spectra contain the same integrated
intensity, however, only the appearance of a clear line feature in the right
spectrum at the expected velocity strongly indicates that the signal is not
due to baseline features but cannot be anything but CO emission.

\subsection{Fitting the CO Line}
\label{sec:fitting}

To extract CO line emission, we perform an automated line fit to each
stacked spectrum. This approach picks out spectral line emission
rather than baseline structure and does not require us to define an
integration window beforehand.

\begin{figure}
\epsscale{1.1} \plotone{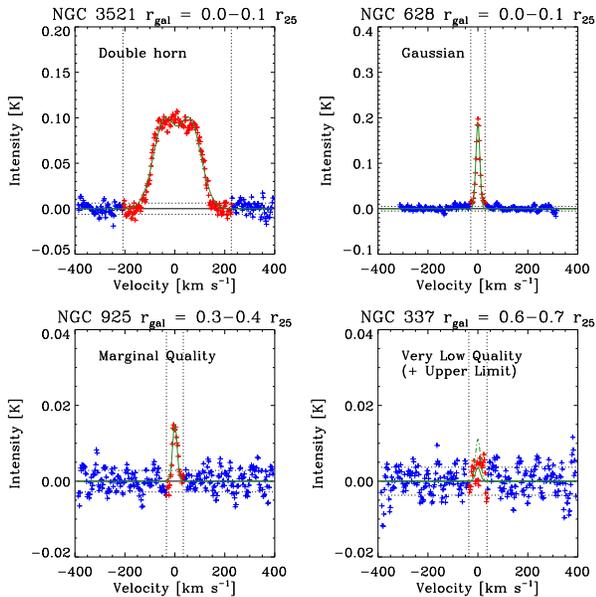}
\caption{\label{f3}Examples of stacked CO spectra with different
  line shapes. The upper row shows spectra with high quality fits and
  the lower row shows spectra with low quality fits. The horizontal
  dotted lines mark the $1\sigma$ rms noise of the stacked spectrum.
  The \emph{upper left} panel shows a broad line approximated by a
  double--horn profile. The \emph{upper right} panel shows a narrow
  line fitted by a Gaussian. The \emph{lower left} panel shows a
  marginal quality fit. The \emph{lower right} panel shows a low quality
  fit together with the associated upper limit (dashed line).}
\end{figure}

In most regions, the line can be well--approximated by a Gaussian
profile with FWHM of $\sim$$15 - 40$ \kmpers\ (Figure~\ref{f3}
upper right panel).  However, in the central regions of some galaxies
the line can be very broad with a flattened or double--horned peak
(Figure~\ref{f3} upper left panel). These profiles often coincide with
central enhancements like bars or molecular rings and are poorly
parametrized by a single Gaussian. Instead we fit a double--horn
profile, a Gaussian scaled by a symmetric second--order polynomial
\citep[functional form from][]{Saintonge2007}.  Based on ``by eye''
inspection, the asymmetry in these profiles is small enough that a
symmetric function is sufficient.

We derive the best fit profile via a non--linear least squares
fit\footnote{We use the IDL procedure \texttt{MPFIT.PRO} from Craig
  Markwardt which performs a Levenberg--Markwardt non--linear least
  squares minimization and is based on the \texttt{MINPACK-1 LMDIF.F}
  algorithm from More et al. (1978)}. We constrain the fit parameters
so that the center of the profile lies within $\pm$50 \kmpers\ of zero
velocity after shifting, the FWHM is larger than 15 \kmpers\ (to avoid
the fit latching onto individual channels), and the amplitude is
positive.  We always carry out a Gaussian fit first and in those cases
where the FWHM exceeds 60 \kmpers\ we switch to a double--horn
profile. We verify by eye that this yields sensible results.

The integral of the fitted profile gives us the integrated CO line
intensity. We derive the uncertainty in this quantity from the noise,
estimated from the signal--free part of the spectrum, and the width
of the profile. It proved useful to define a quality scale for the fit. The
quality is ``high'' where the peak intensity is larger than $5\sigma$
and its integrated intensity is larger than $10$ times its uncertainty.
In cases where the peak intensity is less than $3\sigma$ or the
integrated intensity is less than $5$ times its uncertainty, we do not
trust the fit and instead determine an upper limit. We label cases
that fall between these regimes as ``marginal''. Figure~\ref{f3}
shows examples of our quality measures and line profile fits.

We derive our upper limits integrating over a Gaussian line profile
with FWHM set to 18~\kmpers , the typical FWHM found for high SNR
spectra at $r > 0.5$ $r_{\rm 25}$, and fixed amplitude of 3$\sigma$.

\subsection{Stacking as a Function of Radius}
\label{sec:stack_by_rad}

We present our stacking technique applied to radial bins. In principle
this method allows us to stack spectra across any region. For example,
we could define regions by total gas column, infrared intensity, or
features such as spiral arms and bars. In practice, radius makes
an excellent ordinate. We wish to study the underlying relationship
between CO, \hi, IR, FUV, and \ha\ intensity. Stacking with one of these
quantities as the ordinate would require carefully modeling the biases
involved to measure the underlying relationships. Galactocentric
radius is a well--determined, independent quantity that is also highly
covariant with these other intensities. This yields a dataset with
large dynamic range that is easy to interpret.

Therefore we focus our analysis on data stacked in bins of
galactocentric radius. We average over tilted rings 15\arcsec\
wide, comparable to the angular resolution of our data. This width
corresponds to $\sim$$220$~pc for our nearest targets (3~Mpc)
and $\sim$$1800$~pc for our most distant targets (25~Mpc). We
construct the rings assuming that each galaxy is a thin disk with the
inclination and position angle given in Table~\ref{t1}. To measure
CO with highest sensitivity we construct stacked CO spectra using
the procedure described above and fit those to determine the
integrated CO line intensities. For the other observables --- \hi,
IR, FUV, and \ha\ intensities --- we use two--dimensional maps
of intensity and determine the mean intensity for each tilted ring.

Error bars on the \hi, IR, FUV, and \ha\ intensities show the
$1\sigma$ scatter within that tilted ring, capturing both statistical
noise and deviations from axial symmetry. We estimate the
$1\sigma$ scatter for our CO measurements by integrating
the CO cube over a velocity window that is adjusted for each
line of sight such that it includes all channels of significant
CO emission but at least all channels with velocities within 25
\kmpers\ of the local mean \hi\ velocity. Note that our $1\sigma$
values reflect the scatter in (integrated) intensities of individual
lines of sight inside a ring. They should not be confused with the
uncertainty in the determination of the mean intensity inside a ring
which is typically much smaller.

\section{Results}
\label{sec:results}

\begin{figure*}
\epsscale{1.0} \plotone{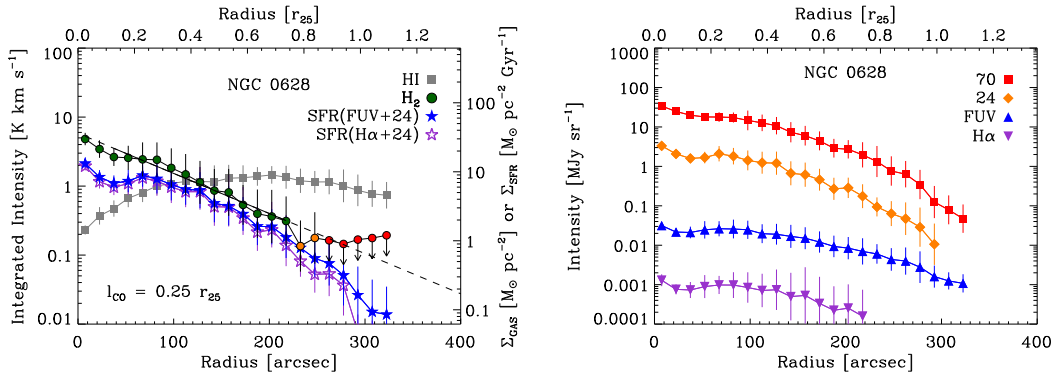}
\caption{Azimuthally averaged intensity in 15\arcsec\ wide tilted
  rings in NGC~628: (\emph{left}) \hi, \htwo , and SFR from 
  FUV$+$24\micron\ and \ha$+$24\micron ; (\emph{right}) 24~\micron ,
  70~\micron , FUV, and \ha\ intensity. Left--hand $y$--axes show
  observed intensities, the right--hand $y$--axis shows the surface
  densities of \hi, \htwo , and SFR projected to face--on values. The
  color of the CO points indicates the significance of the fit to the
  stacked spectrum: high quality in green, marginal quality in orange,
  and upper limits in red. The solid--dashed line shows an exponential
  fit to the radial CO profile with the scale length, $l_{\rm CO}$, printed
  in the lower left corner. Error bars show the $1\sigma$ scatter inside
  each tilted ring. Note that we have scaled the \hi\ intensities (left
  $y$--axis) by a factor of $312.5$ in order to match all profiles in units
  of surface density (see text).\label{f4}}
\end{figure*}

In Figure~\ref{f4} for NGC~628 and in the Appendix for the rest of the
sample, we present stacked radial profiles of integrated CO intensity
along with profiles of \hi , infrared intensity at 24~\micron\ and 70~\micron ,
FUV, and \ha\ intensity. Following Equations~\eqref{eq:sfr_ha24} we
combine \ha\ and 24~\micron\ intensities to estimate the star formation
rate surface density SFR(\ha$+$24) and compare those to SFR(FUV$+$24)
derived from FUV and 24~\micron\ intensities using Equation~\eqref{eq:sfr_fuv24}.

For each galaxy there are two plots: The left panel shows \hi , H$_2$
(from CO), and SFR for both FUV$+$24\micron\ and \ha$+$24\micron.
The right panel shows our SFR tracers --- \ha , FUV, 24~\micron\ 
and 70~\micron\ emission. We present the profiles in both observed
intensity\footnote{In order to have \hi\ and \htwo\ comparable in units
  of mass surface density we scale the observed \hi\ intensity by a
  factor of $312.5$; the ratio of Equation \eqref{eq:hi} and \eqref{eq:h2}.}
(left hand $y$--axis) and units of surface density (right hand $y$--axis
of the left panel) --- $\Sigma_{\rm H2}$, $\Sigma_{\rm HI}$, and
$\Sigma_{\rm SFR}$. Note that we have projected (only) the
surface densities of \hi , \htwo , and SFR to face--on values (i.e., we
corrected for inclination). The observed surface brightnesses are
not corrected for the effect of inclination. Such a correction will just
move each galaxy up and down in lockstep and we find it more useful
to report the observed values. The color of a point in the CO profile
indicates the significance of the fit to the stacked spectrum: green for
high significance, orange for marginal significance, and red for upper
limits, these correspond to $3\sigma$ upper limits on the fitted intensity
(see Section~\ref{sec:fitting}).

\subsection{CO and Star Formation}
\label{sec:sf}

With these azimuthally averaged data we are able to compare CO to
tracers of recent star formation across a large range of \htwo--to--\hi\
ratios. In this subsection, we make empirical comparisons between
measured intensities, examine the relative roles of \htwo\ and \hi\ in
the ``star formation law,'' and investigate the origin of the scatter in
these relations.

\subsubsection{Scaling Relations between CO and IR, FUV, and H$\alpha$}
\label{sec:scalings}

\begin{figure*}
\epsscale{1.0} \plotone{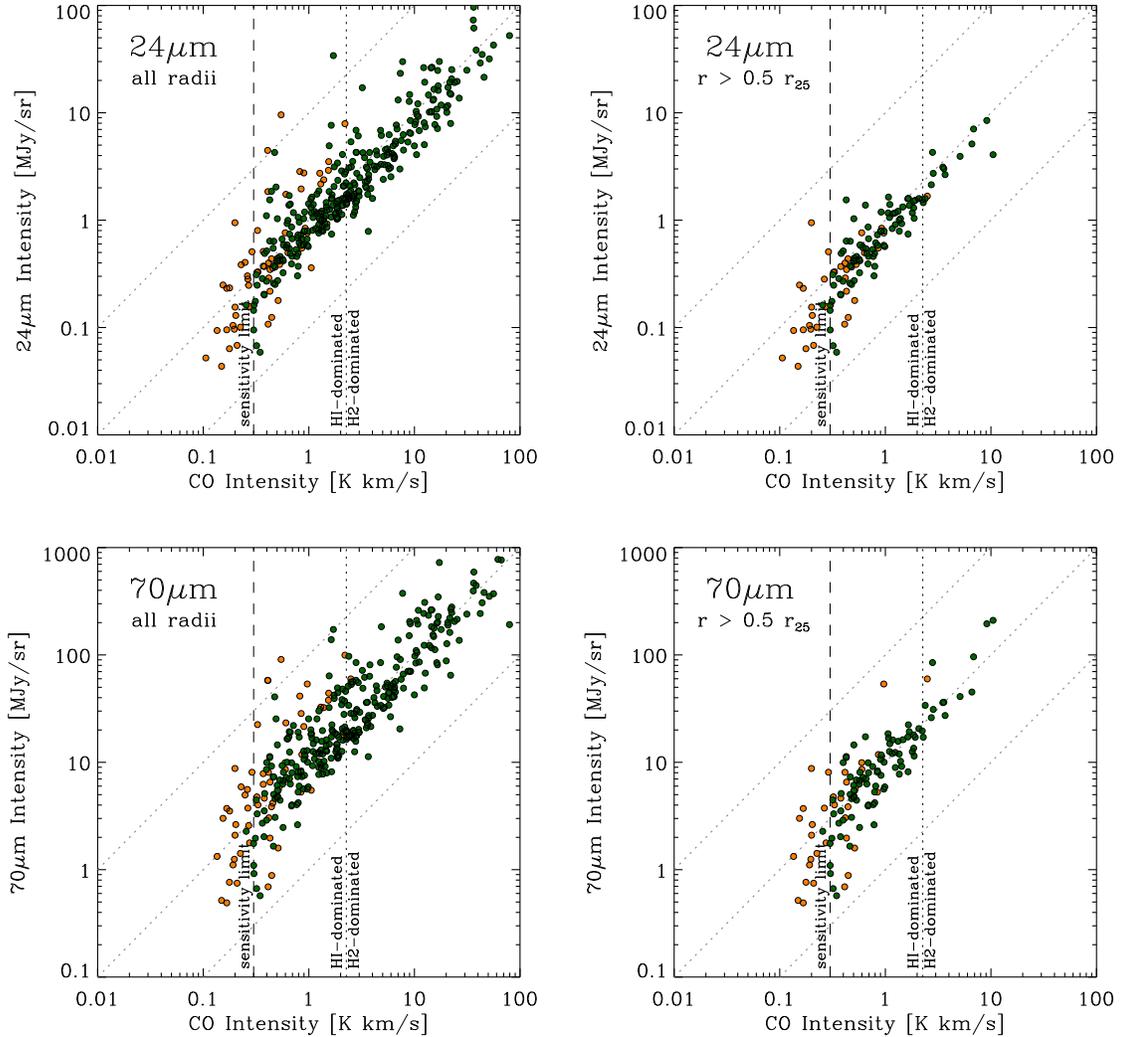}
\caption{Observed IR intensities ($y$--axis) as a function of integrated
  CO\jtwo\ intensity ($x$--axis). Each point corresponds to a stacked
  average in a tilted ring $15$\arcsec\ wide. Green and orange symbols
  indicate CO measurements of high or marginally significance; radial rings
  with only upper limits on the CO intensity (not shown here) are located
  exclusively to the left of the long--dashed line. The top panels shows the
  relation of 24~\micron\ versus CO, the bottom panels shows 70~\micron\ 
  versus CO. The left panels show data for all radii, whereas the right panels
  show only data outside $0.5~r_{\rm 25}$, where the ISM is typically
  \hi --dominated. The short--dashed vertical line indicates a typical CO
  intensity at which $\Sigma_{\rm H2} \sim \Sigma_{\rm HI}$; data to
  the left of this line will usually be \hi--dominated. The diagonal dashed
  lines indicate lines of constant ratios for orientation.\label{f5}}
\end{figure*}

\begin{figure*}
\epsscale{1.0} \plotone{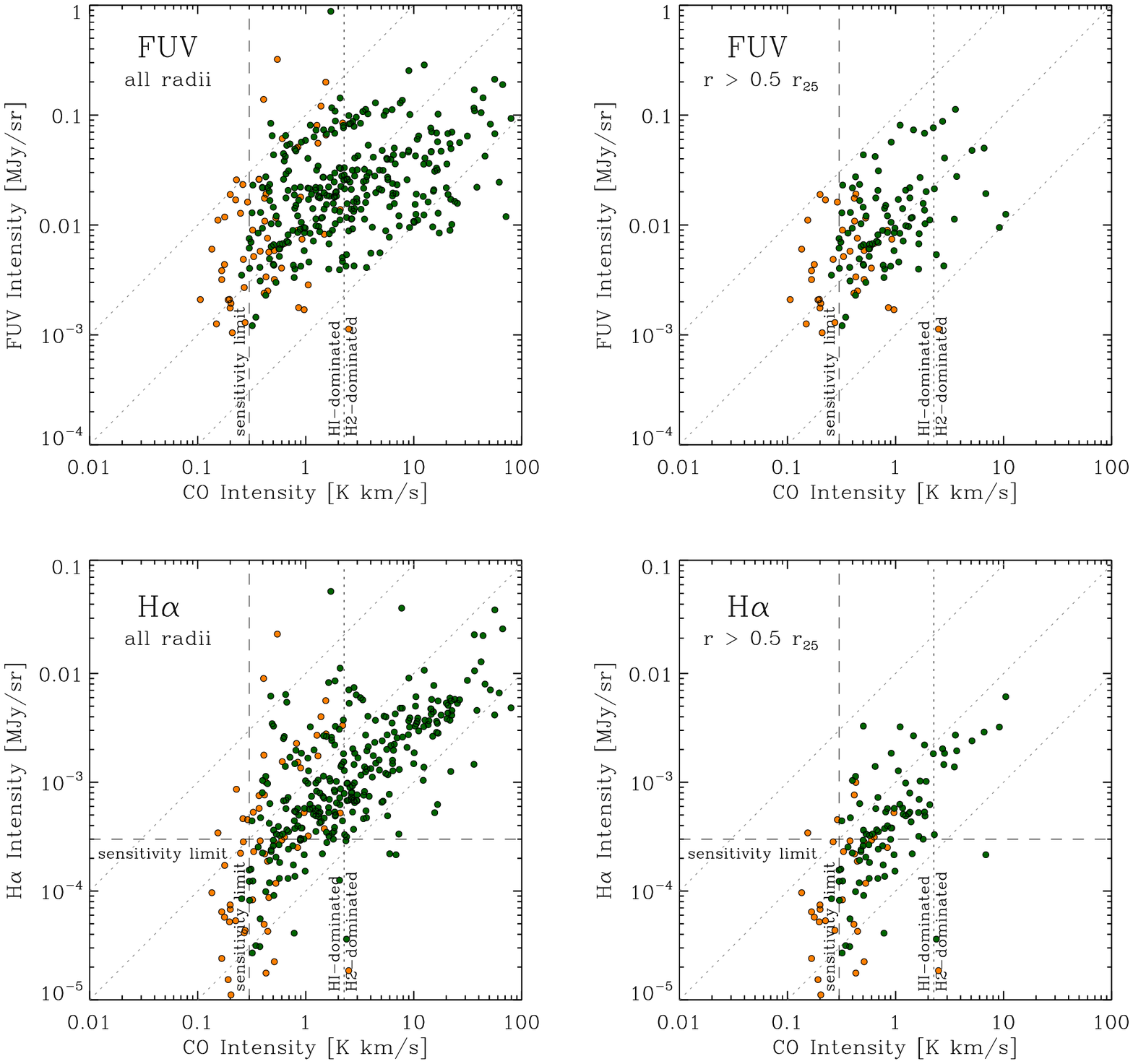}
\caption{A similar plot as Figure~\ref{f5} but this time showing the relation
of FUV versus CO and \ha\ versus CO for our two radial ranges. Note that
\ha\ intensities below a few times $10^{-4}$ \MJypersr\ are affected by data
quality and processing and have to be considered uncertain.\label{f6}}
\end{figure*}

Figures~\ref{f5} \& \ref{f6} show scaling relations between observed
intensities of CO ($x$--axis) and different tracers of recent star formation
($y$--axis). Figure~\ref{f5} shows infrared intensities at 24~\micron\ (top
panels) and 70~\micron\ (bottom panels) and Figure~\ref{f6} shows intensities
of FUV (top panels) and \ha\ (bottom panels). The left hand panels show
the relations for all galaxies and all radii. The panels on the right hand side
show only radii $r > 0.5$ $r_{\rm 25}$. \htwo\ and \hi\ make up roughly
equal parts of the ISM near this radius (see the left panel of Figure~\ref{f13}),
so most of the points in the right hand panels are \hi --dominated. A dotted
vertical line shows an integrated CO intensity of $2.2$~\Kkmpers , which
corresponds to $\Sigma_{\rm H2} \approx 10$~\Msunperpc\ (assuming
$i = 45\arcdeg$), which is about the surface density at which \htwo\ and
\hi\ make up equal parts of the ISM. A dashed vertical line at $I_{\rm CO}
= 0.3$~\Kkmpers\  ($\Sigma_{\rm H2} \approx 1$~\Msunperpc\ for
$i = 45\arcdeg$) shows a conservative sensitivity limit for the whole sample.
Typically our upper limits, which are not displayed in these plots, lie to
the left of this line. They will be systematically higher at large radii when
radial rings partially exceed the coverage of our CO maps.

CO emission correlates tightly with IR emission at 24~\micron\ and 
70~\micron\ (Figure~\ref{f5}) (rank correlation coefficient $r_{\rm corr} 
= 0.9$). The correlation extends over three orders of magnitude in CO
and IR intensities and crosses the \hi --to--\htwo\ transition without
substantial change in slope or normalization. Comparing the left panels
(all radii) and the right panels ($r > 0.5$ $r_{\rm25}$) does not reveal
any significant radial dependence.

CO emission exhibits a weaker correlation with FUV and \ha\
(Figure~\ref{f6}) ($r_{\rm corr} = 0.5 - 0.6$) than with IR emission,
i.e., both the CO--FUV and CO--H$\alpha$ relation show much larger
scatter than the CO--IR relations. The CO--FUV relation displays a
break between \hi-- and \htwo--dominated regimes. The increased
scatter and weaker correlation at least partially reflects the sensitivity
of \ha\ and FUV emission to absorption by dust. In the inner ($r \lesssim
0.5~r_{25}$), more gas rich parts of galaxies dust reprocesses most
\ha\ and FUV emission into IR emission. In this regime CO and FUV
are to first order uncorrelated ($r_{\rm corr} = 0.16$). Outside
$\sim$$0.5~r_{\rm 25}$ the filling factors of dense gas and dust are
lower and FUV is less affected by extinction. The correlation coefficient
between CO and FUV or H$\alpha$ is $r_{\rm corr} = 0.45 - 0.55$ in
this regime, still not as high as for the whole galaxies because of the
limited dynamic range in intensity. In Section~\ref{sec:scatter} we will
see that the weaker correlation of CO with \ha\ and FUV is mostly due
to galaxy--to--galaxy variations, possibly reflecting different star formation
histories, dust abundances, geometries, and potential changes in the
CO--to--H$_2$ conversion factor.

We use the ordinary least squares (OLS) bisector to fit power laws to
each relation. In terms of sensitivity to a given amount of star formation,
the SFR tracers are much more sensitive than our CO maps. To properly
fit a relation between them we thus need to either carefully incorporate
upper limits \citep[e.g., see][]{Blanc2009} or impose a matched sensitivity
cut on the SFR tracer data. We take the latter approach, discarding data
below $0.1$, $1.0$, $10^{-3}$, and $10^{-4}$ \MJypersr\ at 24~\micron,
70~\micron, FUV, and \ha\ intensities after an initial fit\footnote{The result
  is relatively insensitive to the exact choice of sensitivity cut. Varying it by
  a factor of 2 affects the power law index by $\sim$0.05 for the CO--IR
  relations and $\sim$0.15 for CO--FUV or CO--H$\alpha$.}. Graphically,
this removes the flaring towards low intensity just above our sensitivity
cut seen in Figure~\ref{f5} \& \ref{f6}, because we are not sensitive to
a similar flaring towards low CO intensities.

Table~\ref{t2} reports these fits considering all regions with CO
measurements of high significance and SFR tracers above the
sensitivity cut. The main result is that CO emission is consistent
with being linearly proportional to IR emission both at 24~\micron\
and 70~\micron , down to low surface brightness. The best fits
relating CO with FUV or \ha\ emission are also consistent with
a linear slope within the large uncertainties, but power laws are
clearly an inadequate description of those data.

\begin{deluxetable}{lcc}
\tablecolumns{3}
\tablecaption{Relation of IR Emission and SFR Tracers to CO \label{t2}}
\tablehead{\colhead{SFR Tracer} & \colhead{Rank Correlation} & \colhead{Power Law Index}}
\startdata
24~\micron\ & & \\
... all data & $0.90 \pm 0.05$\tablenotemark{a} & $1.0 \pm 0.1$\tablenotemark{b} \\
... $r > 0.5~r_{25}$ & $0.87 \pm 0.10$ & $1.1 \pm 0.1$ \\[2mm]
70~\micron\ & & \\
... all data & $0.87 \pm 0.05$ & $1.0 \pm 0.1$ \\
... $r > 0.5~r_{25}$ & $0.84 \pm 0.10$ & $1.1 \pm 0.2$ \\[2mm]
FUV & & \\
... all data & $0.47 \pm 0.05$ & $0.8 \pm 0.6$ \\
... $r > 0.5~r_{25}$ & $0.46 \pm 0.09$ & $1.1 \pm 0.8$ \\[2mm]
H$\alpha$ & & \\
... all data & $0.63 \pm 0.05$ & $0.9 \pm 0.4$ \\
... $r > 0.5~r_{25}$ & $0.55 \pm 0.10$ & $1.1 \pm 0.6$
\enddata
\tablenotetext{a}{We estimate the uncertainty in the rank correlation
coefficient by taking the correlation coefficient derived from 1,000
random pairwise re--orderings of the data.}
\tablenotetext{b}{The slope quoted here is from the ordinary least
squares bisector with the error estimated from the spread in fitting
$x$ vs. $y$ and $y$ vs. $x$.}
\end{deluxetable}

\subsubsection{Scatter in the Scaling Relations}
\label{sec:scatter}

Each of the observed relations displays significant scatter:
$I_{\rm 24\mu m}/I_{\rm CO}$ has $1\sigma$ scatter of about $0.17$~dex
($\sim$50\%) and $I_{\rm 70\mu m}/I_{\rm CO}$ scatters by $0.24$~dex
($\sim$75\%), whereas $I_{\rm H\alpha}/I_{\rm CO}$ scatters by
$0.34$~dex (a factor of $2.2$) and $I_{\rm FUV}/I_{\rm CO}$ by
$0.55$~dex (a factor of $3.5$). The origin of this scatter is of
astrophysical interest. On small scales,  this scatter arises from
the evolution of individual star--forming regions, which vary
dramatically in their ratios of CO--to--SFR tracers --- leading
to a breakdown of scaling relations when a resolution element
corresponds to an individual region \citep{Schruba2010}. Here
our azimuthal averaging washes out such small--scale variations.
Each point averages over many individual star--forming regions.
However, we do have the ability to distinguish scatter {\em within}
a galaxy from scatter {\em among} galaxies.

To investigate the origin of the observed scatter, we remove
galaxy--to--galaxy variations from the observed relation. We do
so in two ways: First, we fit power--laws relating CO to IR, FUV,
or \ha\ in each galaxy and then adjust all galaxies to have the same
normalization. {\em A priori} we do not know if the galaxy--to--galaxy
variations mainly affect the measurement of our gas tracer ($x$--axis)
or the measurements of our SFR tracers ($y$--axis), therefore, we
repeat the exercise matching normalizations at a fixed value in $y$
and then in $x$. We then compare the scatter among normalizations
to the scatter about the re--normalized relation.

\begin{figure*}
\epsscale{0.95} \plottwo{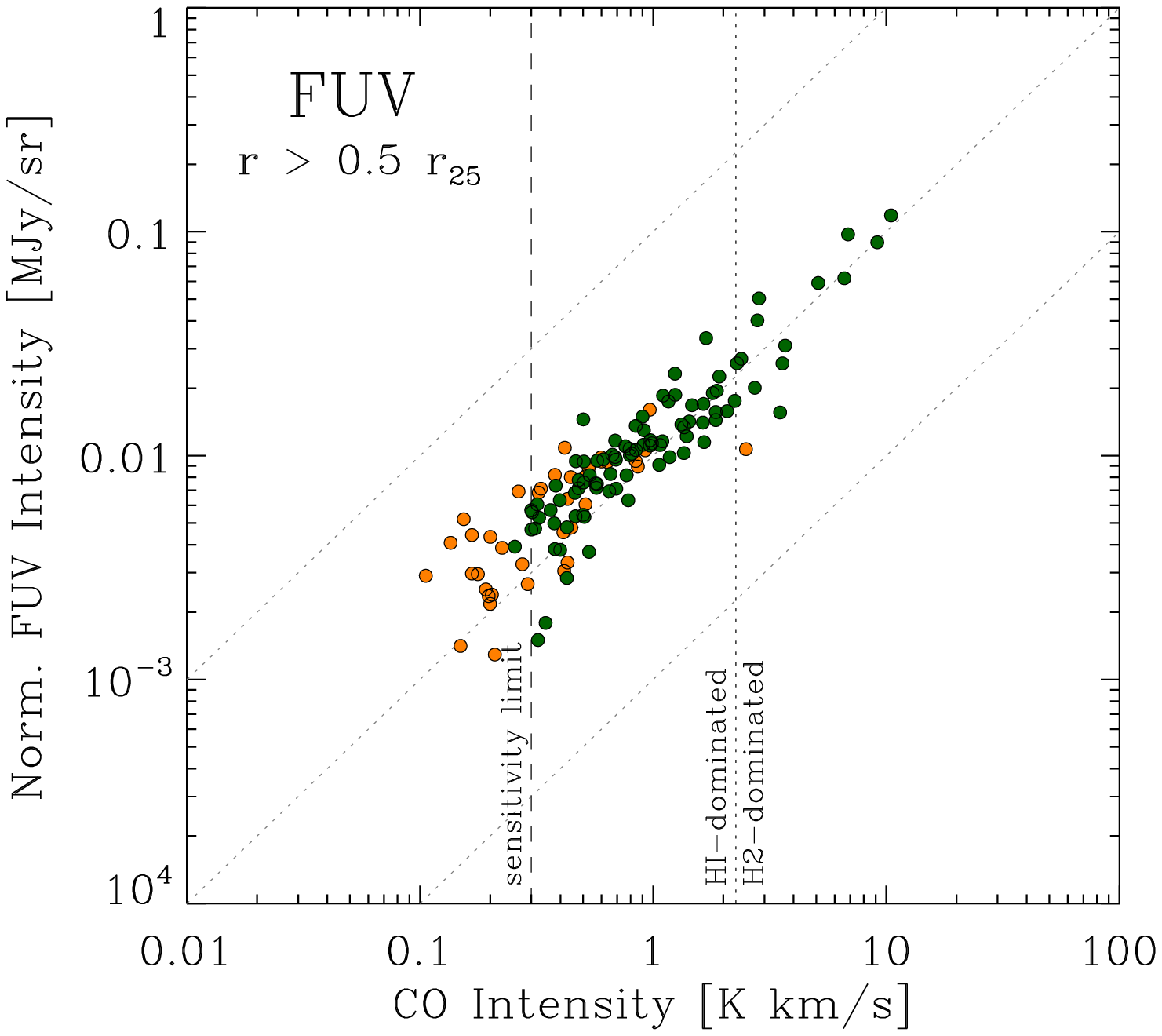}{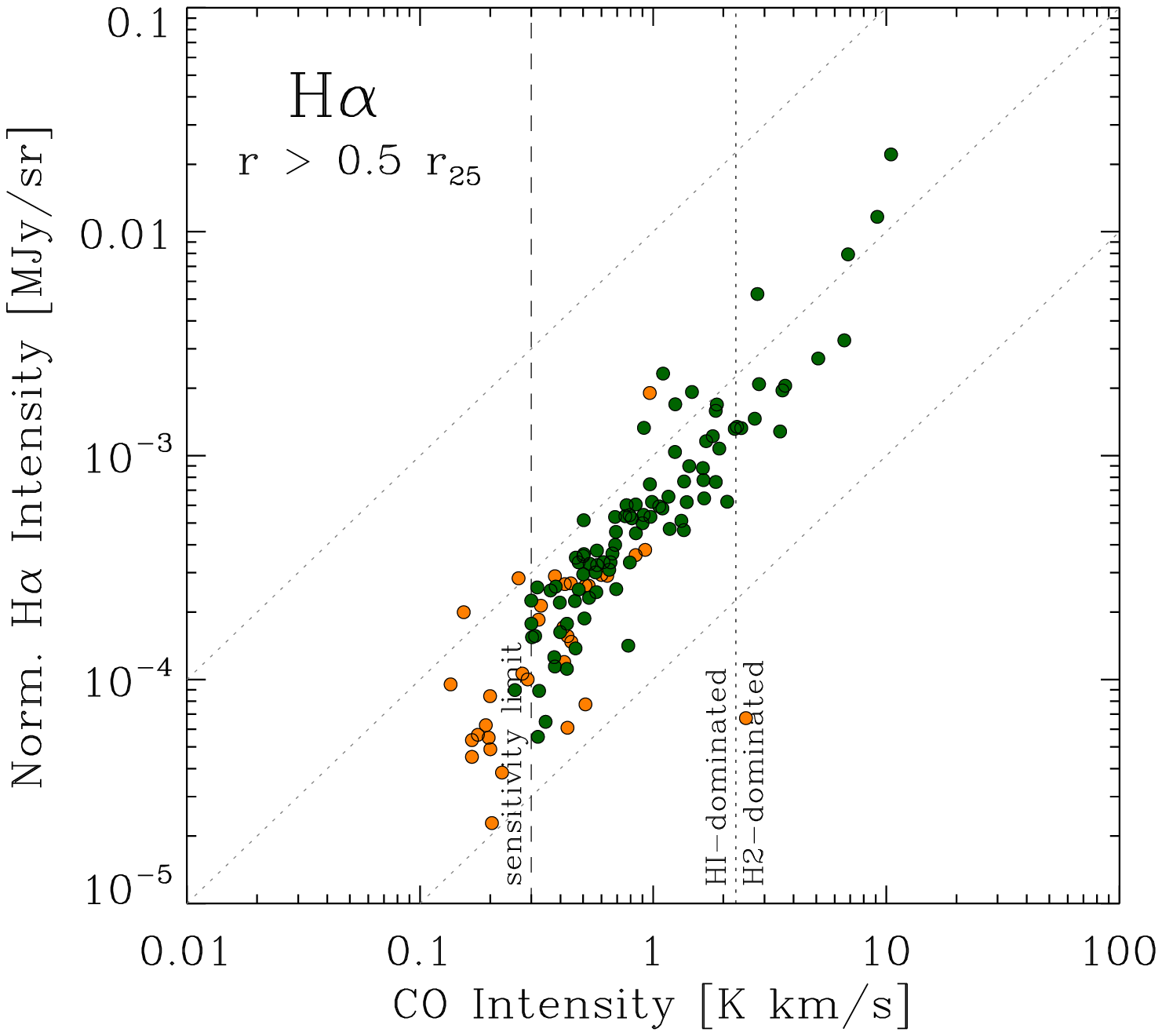}
\caption{Relation between CO intensity and normalized FUV intensity
  (left panel) and normalized \ha\ intensity (right panel) for outer disks,
  $r > 0.5$ $r_{\rm 25}$. For each galaxy we use power law fits to the
  data of high quality and normalize the relation by the FUV or \ha\
  intensity, respectively, which the fit has at 1~\Kkmpers.\label{f7}}
\end{figure*}

We also carry out a more basic test, measuring how the scatter in the
ratio of SFR tracer to CO emission varies both among and within galaxies.
We compare the scatter in the median ratio of SFR tracer to CO,
$\left< I_{\rm 24\mu m}/I_{\rm CO} \right>$, among galaxies to the scatter
in deviations from this median ratio within galaxies, $I_{\rm 24\mu m}/I_{\rm CO}
- \left< I_{\rm 24\mu m}/I_{\rm CO} \right>$.

For both approaches, we find that galaxy--to--galaxy variations
dominate the scatter in the observed relation. Scatter among galaxies
in Figure \ref{f5} \& \ref{f6} is $\sim$2 times larger than the scatter
within individual galaxies. The most striking cases are \ha\ and FUV
emission in the outer parts of galaxies. Figure~\ref{f7} shows that
after normalization, the relation of CO with FUV and \ha\ emission
for radii $r > 0.5$ $r_{\rm 25}$ becomes very strong ($r_{\rm corr} =
0.9$) and nearly linear (power law index of $0.90 \pm 0.05$ for FUV
and $1.1 \pm 0.05$ for \ha). Thus in the outer parts of galaxies the
FUV--to--CO and the \ha--to--CO ratios vary dramatically among
galaxies but are largely fixed inside each galaxy. The scatter appears
driven at least in part by real systematic variations in the ratio of
CO--to--SFR tracers as a function of other galaxy parameters
\citep[see below and][]{Young1996}. The case in the inner parts
of galaxies is more complex because of high dust attenuations
resulting in non--linear CO--FUV and CO--\ha\ relations.

\subsubsection{H$_2$ and Star Formation}
\label{sec:sflaw}

\begin{figure*}
\epsscale{1.0} \plotone{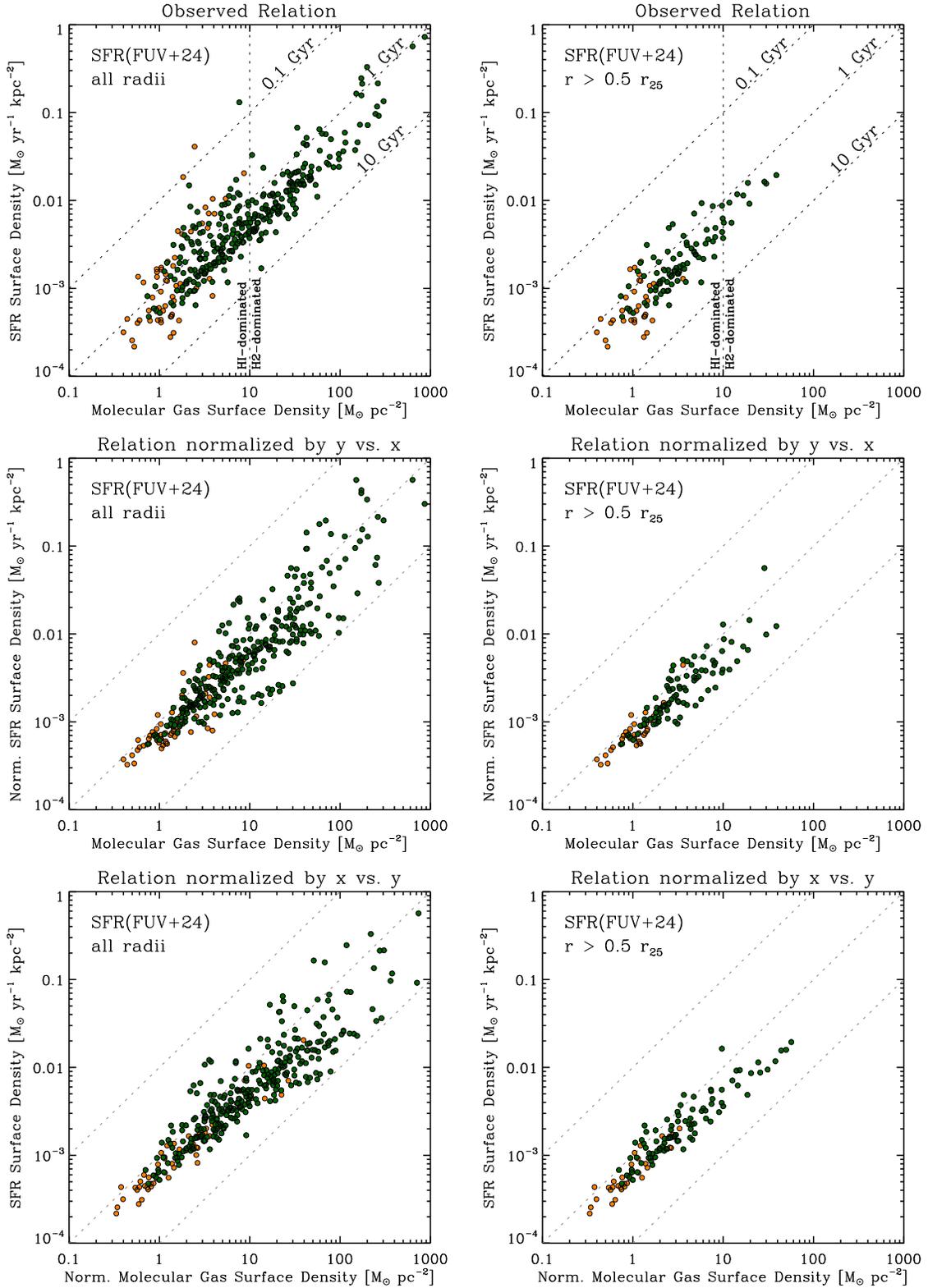}
\caption{Recent star formation rate surface density estimate from
  FUV$+$24\micron\ ($y$--axis) as a function of molecular gas surface
  density ($x$--axis). The left--hand column shows data for all radii,
  whereas the right--hand column shows only data where $r > 0.5~r_{\rm 25}$.
  The upper row presents the basic relation, which is approximately linear
  with average H$_2$ depletion time $\sim$$2.0$~Gyr (dashed lines). The
  middle and bottom rows show results after we fit and remove galaxy--to--galaxy
  variations. These galaxy-to-galaxy variations are $\sim$2 times larger than
  internal variations, suggesting individual galaxies each obey well--defined,
  though offset, relations. \label{f8}}
\end{figure*}

\begin{figure*}
\epsscale{1.0} \plotone{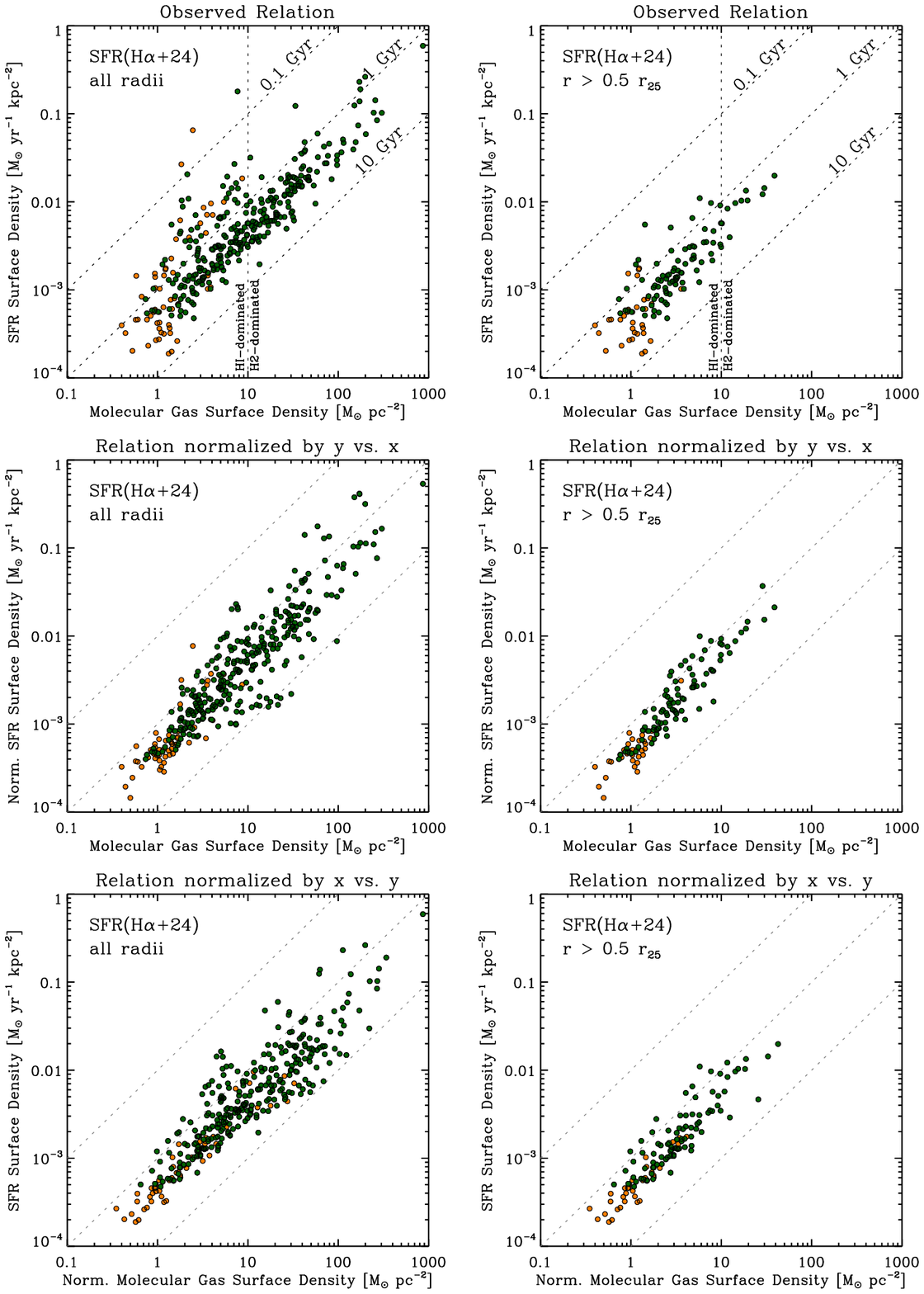}
\caption{The same plot as Figure~\ref{f8} but this time using \ha$+$24\micron\
  to estimate $\Sigma_{\rm SFR}$. \label{f9}}
\end{figure*}

In Figure~\ref{f8} we combine FUV and 24~\micron\ intensities
to estimate $\Sigma_{\rm SFR}$, which we plot as a function of
$\Sigma_{\rm H2}$. In Figure~\ref{f9} we instead combine \ha\
and 24~\micron\  to estimate $\Sigma_{\rm SFR}$. In both Figures,
the left hand panels show data for all radii, while the right hand panels
show only rings with $r > 0.5$ $r_{\rm 25}$, where the ISM is mostly
\hi . As in Figure \ref{f5} \& \ref{f6}, a vertical dotted line shows
$\Sigma_{\rm H2} \approx 10$ \Msunperpc , a typical \htwo\ surface
density where the ISM consists of equal parts \hi\ and \htwo .

In agreement with \citet{Bigiel2008,Bigiel2011}, we observe an
approximately linear scaling of $\Sigma_{\rm SFR}$ and $\Sigma_{\rm H2}$
in regions that are dominated by molecular gas ($\Sigma_{\rm H2} \gtrsim 10$
\Msunperpc). For Figure~\ref{f8} this agreement does not come as a
surprise because we use many of the same data and a similar approach
(FUV$+$24\micron) to estimate recent SFR. The new results here are that
(a) this trend {\em continues without significant changes} down to very low
$\Sigma_{\rm H2}$, including regions strongly dominated by atomic gas,
and (b) we find the same trend in Figure~\ref{f9} using SFR(\ha$+$24\micron).

The rank correlation coefficient relating $\Sigma_{\rm H2}$ and
$\Sigma_{\rm SFR}$ for all data with at least a marginal CO measurement
is $r_{\rm corr} \approx 0.85$. The scatter about a linear relation is
$0.25$~dex (prior to any normalization). Both numbers resemble those
derived for the CO--24~\micron\ relation in Section~\ref{sec:scalings}
because 24~\micron\ emission drives our hybrid SFR tracer over most
of the area. In detail, the fractional contribution of 24~\micron\ to SFR
varies with radius and choice of hybrid tracer. Generally speaking:
(a) the larger the radius the larger the contribution of the unobscured
term, and (b) \ha\ contributes fractionally more than FUV to the hybrid.

An OLS bisector fit yields a roughly linear slope and a molecular
depletion time, $\tau_{\rm dep} = \Sigma_{\rm H2} / \Sigma_{\rm SFR}$,
of $\sim$$1.8$~Gyr (including a factor $1.36$ to account for heavy 
elements; see Table~\ref{t3} for fit parameters). This is slightly lower 
than $\tau_{\rm dep} = 2$~Gyr found by \citet{Bigiel2008} and 
\citet{Leroy2008} for a subset of the data analyzed here and
$\tau_{\rm dep} = 2.35$~Gyr recently found by \citet{Bigiel2011} for
a sample that is similar to the one analyzed here. We include (a) more
starburst galaxies and (b) more low mass, low metallicity spiral galaxies
that \citet{Bigiel2008} and \citet{Leroy2008} excluded from their CO
analysis. Both dwarfs and starbursts have shorter $\tau_{\rm dep}$
than large spirals \citep[see below and][]{Kennicutt1998, Gao2004,
Leroy2008, Daddi2010}. Moreover, our radial profiles weight these
small galaxies more heavily than the pixel sampling used by
\citet{Bigiel2008,Bigiel2011}.

As with the scaling relations between CO and tracers of recent star
formation (Figure~\ref{f5} \& \ref{f6}), the H$_2$--SFR relation exhibits
significant scatter. We perform the same procedure to isolate
galaxy--to--galaxy variations from scatter within galaxies and again
find the scatter in the main relation (upper panels in Figure~\ref{f8} \&
\ref{f9}) dominated by galaxy--to--galaxy variations. We plot the relations
after normalization in $x$ and $y$ in the middle and bottom panels of
Figure~\ref{f8} \& \ref{f9}. Once galaxy--to--galaxy scatter is removed,
there is a remarkably tight, uniform linear relation linking molecular gas
and star formation across almost three and a half orders of magnitude.

\begin{figure}
\epsscale{1.0} \plotone{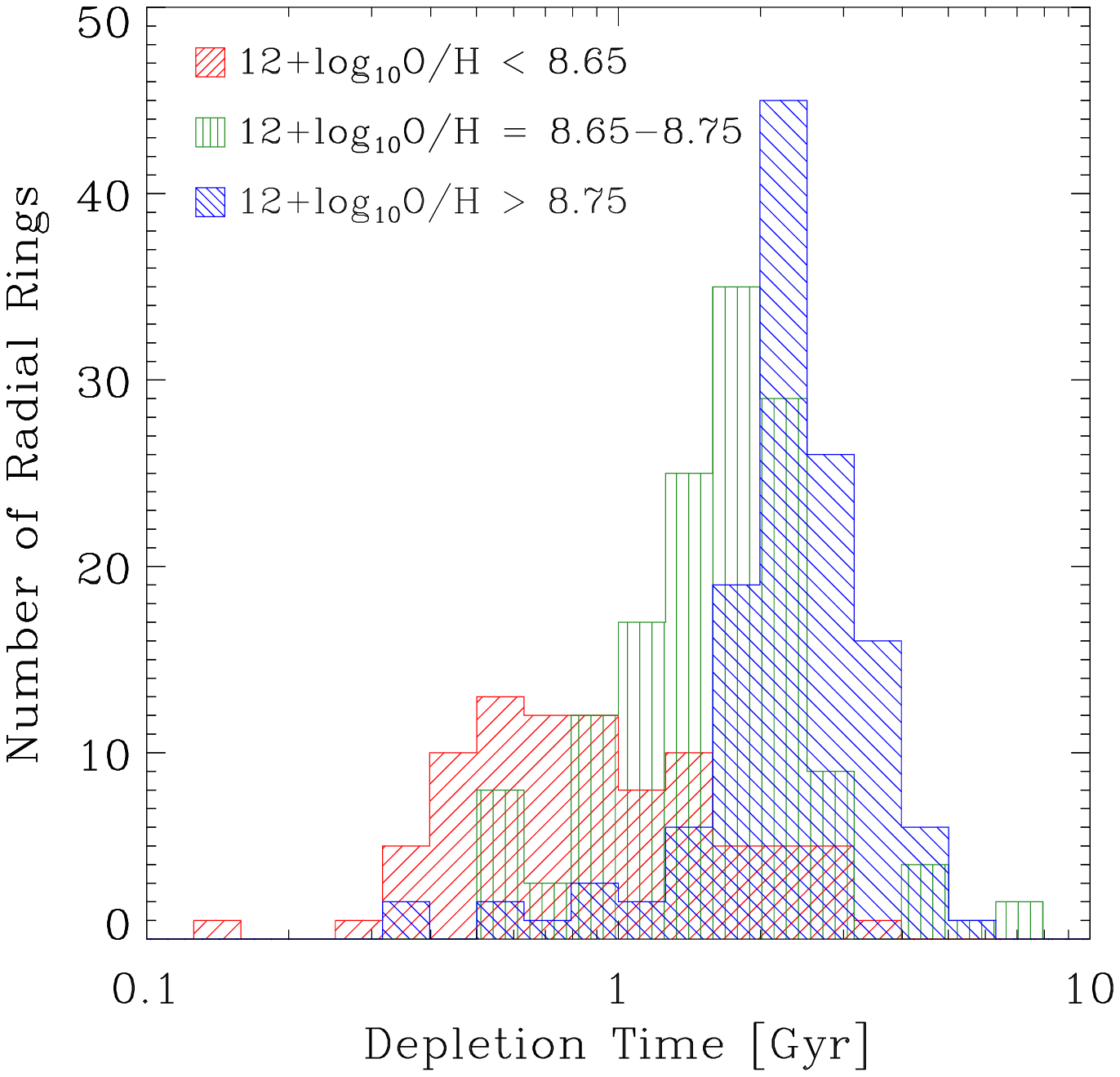}
\caption{Molecular gas depletion time, $\tau_{\rm dep} = \Sigma_{\rm H2}
  /  \Sigma_{\rm SFR}$, as function of metallicity. The depletion times
  shown here correspond to the ratio of the data shown in Figure~\ref{f8}
  including regions with CO measurements of marginal significance. The
  median average $\tau_{\rm dep}$ changes systematically with metallicity:
  for the low metallicity bin $\sim$$0.8$~Gyr, for the intermediate metallicity
  bin $\sim$$1.7$~Gyr, and for the high metallicity bin $\sim$$2.4$~Gyr.
  \label{f10}}
\end{figure}

What drives this galaxy--to--galaxy variation? In both the H$_2$--SFR
and observed intensity relations a large part of the scatter comes from
a sub--population of less massive, less metal--rich galaxies that exhibit
high SFR--to--CO ratios. Figure~\ref{f10} shows the metallicity dependence
of the molecular depletion time, $\tau_{\rm dep}$. Massive spiral galaxies
with high metallicities have considerably longer (median averaged)
depletion times: $\sim$$2.4$~Gyr for systems with metallicities of $12+\log
{\rm O/H} > 8.75$ and $\sim$$1.7$~Gyr for $12+\log {\rm O/H} = 8.65 - 8.75$,
while smaller galaxies with lower metallicities have systematically shorter
depletion times: $\sim$$0.8$~Gyr for systems with $12+\log {\rm O/H} <
8.65$\footnote{The absolute metallicity values that we quote should
not be overemphasized, but the relative ordering of galaxies is fairly
secure. See \citet{Moustakas2010} for more details.}. We find a similar
trend for the scaling of $\tau_{\rm dep}$ with the maximal rotation velocity,
$v_{\rm flat}$, that low mass systems with $v_{\rm flat} \lesssim 140$
\kmpers\ have $\tau_{\rm dep} < 1$~Gyr. These low mass, low metallicity
systems are atomic dominated ($\Sigma_{\rm HI} \gtrsim \Sigma_{\rm H2}$)
for most of their radii and show up prominently in the upper left panel of
Figure~\ref{f8} \& \ref{f9} as the data points offset to shorter depletion times.
\citet{Leroy2008} and \citet{Bigiel2008} labeled these galaxies
``\hi--dominated'' and did not consider them in their H$_2$--SFR analysis.

\subsubsection{SFR, HI, and H$_2$}
\label{sec:gastype}

\begin{figure*}
\epsscale{1} \plotone{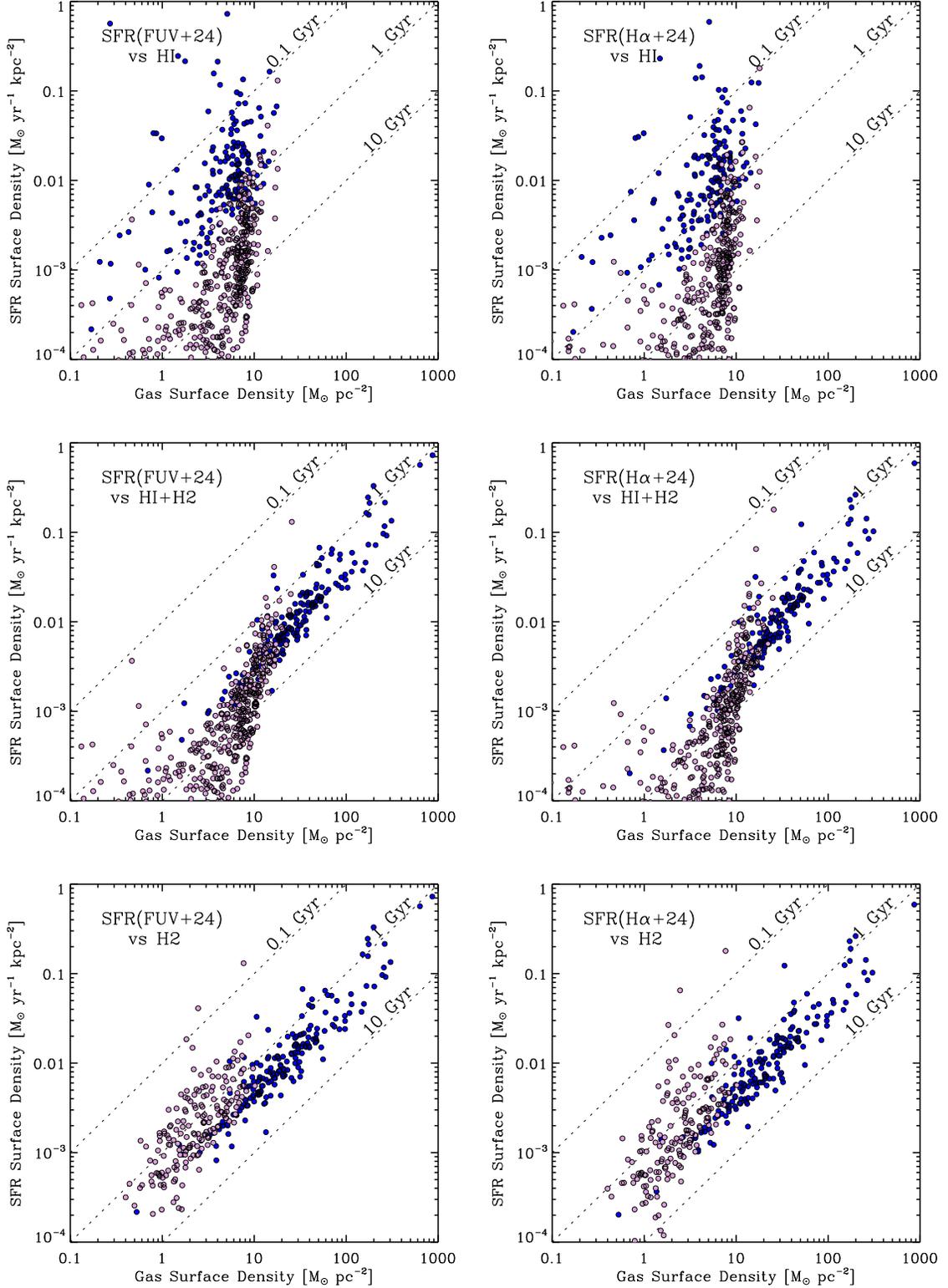}
\caption{$\Sigma_{\rm SFR}$ ($y$--axis) from FUV$+$24$\mu$m ({\em left})
  and \ha$+$24$\mu$m ({\em right}) as a function of different gas phases: \hi\
  alone ({\em top}), \hi+\htwo ({\em middle}), and H$_2$ alone ({\em bottom}).
  Each point in these diagrams represents a radial average in a given galaxy.
  Regions that are \htwo --dominated are plotted with dark blue symbols,
  regions that are \hi --dominated in light red symbols. The bottom panels
  show only regions with at least marginal CO signal, while the top and middle
  panels show also regions where we determined only an upper limit on the
  molecular content. Whereas SFR is not correlated to \hi\ (in the inner parts
  of galaxy disks), it correlates with H$_2$ and total gas. The scaling with H$_2$
  is uniform and linear for all regimes; the scaling with total gas exhibits a change
  in slope at the transition between \hi-- and \htwo--dominated environments.
  \label{f11}}
\end{figure*}

\begin{deluxetable}{lcc}
\tablecolumns{3}
\tablecaption{Relation of SFR to Different Gas Phases \label{t3}}
\tablehead{\colhead{Gas Phase} & \colhead{Rank Correlation} & \colhead{Power Law Index}}
\startdata
\cutinhead{SFR(FUV$+$24) vs. Gas Phase}
\hi\ & & \\
... all data & $0.23 \pm 0.05$\tablenotemark{a} & \nodata \\
... \hi--dominated & $0.52 \pm 0.06$ & \nodata \\
... \htwo--dominated & $0.41 \pm 0.08$ & \nodata \\
\htwo\ & & \\
... all data & $0.88 \pm 0.05$ & $1.0 \pm 0.1$\tablenotemark{a} \\
... \hi--dominated & $0.75 \pm 0.07$ & $1.3 \pm 0.5$ \\
... \htwo--dominated & $0.91 \pm 0.07$ & $1.1 \pm 0.1$ \\
\hi+\htwo\ & & \\
... all data & $0.90 \pm 0.05$ & $1.6 \pm 0.2$ \\
... \hi--dominated & $0.78 \pm 0.06$ & $2.2 \pm 0.6$ \\
... \htwo--dominated & $0.92 \pm 0.07$ & $1.2 \pm 0.1$ \\
\cutinhead{SFR(H$\alpha$$+$24) vs. Gas Phase}
\hi\ & & \\
... all data & $0.21 \pm 0.05$\tablenotemark{a} & \nodata \\
... \hi--dominated & $0.54 \pm 0.06$ & \nodata \\
... \htwo--dominated & $0.44 \pm 0.08$ & \nodata \\
\htwo\ & & \\
... all data & $0.85 \pm 0.05$ & $1.0 \pm 0.1$\tablenotemark{a} \\
... \hi--dominated & $0.70 \pm 0.07$ & $1.5 \pm 0.7$ \\
... \htwo--dominated & $0.91 \pm 0.08$ & $1.1 \pm 0.1$ \\
\hi+\htwo\ & & \\
... all data & $0.90 \pm 0.05$ & $1.7 \pm 0.2$ \\
... \hi--dominated & $0.79 \pm 0.06$ & $2.5 \pm 0.8$ \\
... \htwo--dominated & $0.92 \pm 0.08$ & $1.3 \pm 0.1$
\enddata
\tablenotetext{a}{Estimation of uncertainties equal to Table~\ref{t2}.}
\end{deluxetable}

The question of which gas component --- \hi , \htwo , or total gas ---
correlates best with recent star formation has received significant
attention. Phrased this way, the question is not particularly well
posed: the total gas surface density and the molecular gas fraction
are closely related so that the different gas surface densities are
not independent quantities. Therefore we do not necessarily
expect a ``best'' correlation, only different functional forms. Still,
it is illustrative to see how recent star formation relates to each
gas tracer. Table~\ref{t3} lists $r_{\rm corr}$ and the power law
index from an OLS bisector fit between each component and
Figure~\ref{f11} shows plots for each of the three gas phases and
our two SFR prescriptions. The rank correlation coefficient and the
power law fits are determined for regions where we have at least
a marginal CO measurement. If we restrict our analysis to high
significance CO measurements, we obtain the same results within
the uncertainties.

The table and figure show that recent star formation rate tracers
rank--correlate approximately equally well with \htwo\ and \hi+\htwo\
both across all surface densities and separately in the \hi-- and
\htwo--dominated regimes. \hi\ does not correlate significantly with
star formation in the \htwo--dominated inner parts of galaxy disks,
though the correlation between \hi\ and recent star formation
becomes stronger in the \hi--dominated outer parts of galaxies
\citep[and further increases at even larger radii; see][]{Bigiel2010b}.

The rank correlation is a non--parametric measure of how well
the relative ordering of two data sets align. A high rank correlation
coefficient implies a monotonic relationship but not a fixed
functional form law. For total gas, the power law index relating gas
and recent star formation depends fairly strongly on the subset of
data used. If we focus on the regions where $\Sigma_{\rm HI} >
\Sigma_{\rm H2}$, the best--fit power law relating total gas and
recent star formation (from FUV$+$24) has an index of $2.2 \pm
0.6$. Where $\Sigma_{\rm HI} < \Sigma_{\rm H2}$, the index is
much shallower, $1.2 \pm 0.1$. By contrast, the power law relating
H$_2$ to recent star formation varies less across regimes, from
$1.3 \pm 0.5$ where $\Sigma_{\rm HI} > \Sigma_{\rm H2}$ to $1.1
\pm 0.1$ where $\Sigma_{\rm HI} < \Sigma_{\rm H2}$. Both of these
agree within the uncertainties with the fit to all data, which has
slope $1.0 \pm 0.1$. The slight steepening of the relation in the
outer disks ($\Sigma_{\rm HI} > \Sigma_{\rm H2}$) is driven by the
interplay of two effects: the rather small dynamic range in gas surface
densities and an increased dispersion in SFR--to--H$_2$ ratios
driven by the low mass galaxies; those have high SFR--to--H$_2$
ratios and contribute mostly to the \hi--dominated subset. The
qualitative picture of a break in the total gas--SFR relation but a
continuous H$_2$--SFR relation remains unchanged if SFR(\ha$+$24)
is considered, though there are small changes in the exact numbers
(see the left and right panels in Figure~\ref{f11}).

Our conclusions thus match those of \citet{Bigiel2008}: a single
power law appears to be sufficient to relate $\Sigma_{\rm H2}$
and $\Sigma_{\rm SFR}$ whereas the relationship between
$\Sigma_{\rm HI+H2}$ and $\Sigma_{\rm SFR}$ varies
systematically depending on the subset of data used. Because
we have a dataset that includes significant CO measurements
where $\Sigma_{\rm HI} > \Sigma_{\rm H2}$ we can extend
these findings. First, total gas and H$_2$ are equally well
rank--correlated with recent star formation in all regimes and
this correlation is always stronger than the correlation of recent
star formation with \hi . Second, the H$_2$--SFR scaling relation
extends smoothly into the regime where $\Sigma_{\rm HI} >
\Sigma_{\rm H2}$ whereas the total gas--SFR relation does not.
Third, the result is independent of our two star formation rate
tracers SFR(FUV$+$24) and SFR(\ha$+$24).

\subsubsection{Discussion of CO--SFR Scaling Relations}

{\em Empirical Results:} IR brightness at both 24~\micron\ and
70~\micron\ correlates strongly with CO intensity over $\sim$3
orders of magnitude. Across this range, there is a nearly fixed
ratio of CO to IR emission. FUV and \ha\ emission show little
or no correlation with CO intensity in the inner parts of galaxies,
presumably due to extinction, but are found to correlate well with
CO in the outer parts of galaxies after galaxy--to--galaxy scatter
is removed.

A result of these empirical scaling relations is that the ratio of recent
star formation rate, traced either by combining FUV and 24~\micron\
or \ha\ and 24~\micron\ intensities, to molecular gas, traced by CO
emission, does not vary strongly between the \hi--dominated and
\htwo--dominated ISM. This result is driven largely by the tight
observed correlation between CO and 24~\micron\ emission.
The tight relation between CO and 70~\micron\ emission suggests
that the CO--IR relation actually holds for a larger range of mid--IR
intensities. The tightening of the CO--FUV and CO--\ha\ relation in
the outer part of galaxies (after removing galaxy--to--galaxy scatter)
reinforce the idea of a linear relation extending to large radii. These
tight correlations are consistent with the conclusion of \citet{Leroy2008}
that there is only weak variation in the SFR per unit molecular gas
mass with local environment.

{\em Galaxy--to--Galaxy Scatter:} Each of the correlations we observe
has significant internal scatter. Breaking this apart into scatter among
galaxies and scatter within galaxies, we observe that in every case
scatter among galaxies drives the overall scatter in the observed
correlation. This ``scatter'' among galaxies is not random; less massive,
less metal--rich galaxies exhibit a higher ratio of SFR tracer to CO emission
\citep[see Figure~\ref{f10} and][]{Young1996}. There are two straightforward
physical interpretations for this. The efficiency of star formation from H$_2$
gas may be genuinely higher in these systems, a view advocated by
\citet{Gardan2007} and \citet{Gratier2010}. Alternatively, CO emission
may be depressed relative to the true amount of H$_2$ mass due to
changes in the dust abundance. Low mass systems often have lower
metallicities and correspondingly less dust, which is required to shield
CO \citep[e.g.,][]{Maloney1988, Bolatto1999, Glover2010, Wolfire2010}.
A precise calibration of the CO--to--H$_2$ conversion factor as a
function of metallicity is still lacking, so it is not possible at present
to robustly distinguish between these two scenarios.

After removing these galaxy--to--galaxy variations we find a series of
extraordinarily tight relationships between CO and tracers of recent
star formation. The most striking --- and puzzling --- example of this
is the emergence of a tight correlation between CO and FUV emission
in the outer parts of galaxies ($r > 0.5~r_{25}$). This is puzzling because
one would expect the galaxy to be mostly causally disconnected over
the timescales predominantly traced by FUV emission --- $20-30$~Myr
compared to a dynamical (orbital) time of a few $100$~Myr. Yet,
somehow the differences between galaxies affect the CO--to--FUV ratio
much more than the differences between the widely separated rings
represented by our data points. Galaxy--wide variations in metallicity
and dust abundance probably offer the best explanation for this. These
may propagate into variations in the CO--to--H$_2$ conversion factor
and the average dust extinction. The latter may lead to scatter in
estimates of SFR and both will affect $\tau_{\rm dep}$. An alternative
explanation is that external processes, which affect the whole galaxy,
play a large role in setting the star formation rate on timescales traced
by FUV emission.

These strong galaxy--to--galaxy variations partially explain the
unexpected lack of correlation between CO emission and recent SFR
observed by \citet{Kennicutt1998}. They averaged across whole
galaxies and in doing so conceivably lost the strong internal relations
that we observe but preserved the large galaxy--to--galaxy variations
that offset internal relations. The result will be an apparent lack of
correlation in galaxy--averaged data that obscures the strong internal
relationship. Whether there is in fact a weaker relationship between
H$_2$ and SFR in galaxy--integrated measurements than inside
galaxies depends on whether the suggested variations in star
formation efficiency are real or a product of a varying CO--to--H$_2$
conversion factor.

{\em (The Lack of) A Molecular Star Formation Law:} With improved
sensitivity to CO emission we now clearly see nearly linear relations
between CO and tracers of recent star formation rate spanning from
the \htwo--dominated to \hi--dominated parts of galaxies. Note that
the relations will likely depart from these scaling relations if regions
of high surface densities ($\Sigma_{\rm H2} > 100$~\Msunperpc) or
starburst galaxies are considered \citep[e.g.,][]{Kennicutt1998,
Daddi2010, Genzel2010}. The lack of strong variations in the scaling
between these two quantities in the ``non--starburst regime'' reinforces
that molecular gas is the key prerequisite for star formation. A nearly
linear correlation over this whole range can also be restated as the
absence of a strong relationship between the ratio $\Sigma_{\rm SFR}
/ \Sigma_{\rm H2}$ and $\Sigma_{\rm H2}$. This implies that
$\Sigma_{\rm H2}$ averaged over a large area is not a key
environmental quantity for star formation because it does not affect
the rate of star formation per unit molecular gas. Apparently the global
amount of H$_2$ directly sets the global amount of star formation but
the surface density of H$_2$ does not affect how quickly molecular gas
is converted to stars.

By contrast, the host galaxy does appear to affect the ratio of star
formation rate to at least CO intensity. This indicates important
environmental variations but they are not closely linked to surface
density. In this sense, Figure~\ref{f8} \& \ref{f9} offer a counterargument
against the idea of a star formation ``law'' in which gas surface density
alone sets the star formation rate. Instead, over the disks of normal galaxies,
we see star formation governed by two processes: (a) the formation of
stars in molecular gas, which varies mildly from galaxy to galaxy but
appears largely fixed inside a galaxy, and (b) the conversion of \hi\ to
\htwo, which does exhibit a strong dependence on environment inside
a galaxy, including a strong dependence on surface density (Section
\ref{sec:htwo}). In the second part of this paper we will look at this second
process by measuring variations in the \htwo --\hi\ balance as a function
of gas surface density and radius.

{\em Systematic Effects:} We have interpreted the observed scaling
relations in terms of a relationship between molecular gas and recent
star formation rate. \citet{Leroy2008} demonstrated that azimuthally
averaged profiles of the FUV$+$24\micron\ combination that we use
here match those of several commonly used star formation rate tracers
with $\sim$50\% scatter. Here we have shown that using a combination
of \ha$+$24\micron\ to determine the SFR leads to indistinguishable
results (see Leroy et al. 2011, in prep. for more discussion). We now
discuss several systematic effects that may affect the translation from
observables to inferred quantities.

The most serious worry is that the IR intensity, which drives the
correlations, is acting as a tracer of dust abundance and not recent
star formation. Gas and dust are observed to be well mixed in the ISM,
so in the extreme, this would result in plotting gas against gas times
some scaling factor (the dust--to--gas ratio). A more subtle version of
the same concern is that CO emission is primarily a function of dust
shielding against dissociating UV radiation. If there are large
variations in the abundance of dust in the ISM then it may be likely
that dust emission and CO emission emerge from the same regions
because that is where CO can form and evade dissociation.

A few considerations suggest that the 24~\micron\ and the
70~\micron\ emission are not primarily tracing dust abundance. First,
over whole galaxies, monochromatic IR emission at 24~\micron\ and
70~\micron\ does track the SFR \citep[with some important variations
  among types of galaxy; see][]{Calzetti2010}. Second, we observe a
linear correlation with CO emission and not with overall gas column,
which one might expect for a dust tracer. Third, both the 24~\micron\
and 70~\micron\ bands are well towards the blue side of the peak of
the IR SED for dust mixed with non star--forming gas \citep{Boulanger1996}
and so are not likely to be direct tracers of the dust optical depth (mass).
Still, a thorough investigation of the interplay between dust abundance,
IR emission, and star formation is needed to place SFR tracers involving
IR emission on firmer physical footing.

A less severe worry is that using dust and FUV emission makes us
sensitive to an old stellar component that might not have formed locally.
Our targets are all actively star--forming systems, so old here means
mainly old relative to H$\alpha$ emission \citep[$\tau \sim 4$~Myr;][]
{McKee1997}. The appropriate timescale to use when relating star
formation and gas is ambiguous. When studying an individual region,
it may be desirable to use a tracer with the shortest possible time
sensitivity. Averaging over large parts of galaxies, one is implicitly
trying to get at the equilibrium relation. Therefore a tracer with a
somewhat longer timescale sensitivity may actually be desirable.
The typical $20 - 30$~Myr timescale \citep{Salim2007} over which
most UV emission (and dust heating from B stars) occurs is well
matched to current estimates for the lifetimes of giant molecular clouds
\citep{Kawamura2009}. This makes for a fairly symmetric measurement
--- with the spatial and time scales of the two axes matched --- though
one is comparing recent star formation with the material of future
star--forming regions.

The fact that SFRs derived from FUV$+$24\micron\ and
\ha$+$24\micron\ are essentially indistinguishable indicates that
the distinction between the time scale probed by \ha\ ($\sim$4~Myr)
and FUV ($20-30$~Myr) is not important to this study, probably
because of the large spatial scales considered by our azimuthal
averages.

There may also be systematic biases in our inferred $\Sigma_{\rm
 H2}$. We have already discussed the dependence of \xco\ on
metallicity as a possible explanation for the high SFR--to--CO ratio
observed in lower--mass galaxies. \xco\ certainly depends on
metallicity. Current best estimates imply a non--linear relationship,
with \xco\ sharply increasing below $12 + \log_{10} {\rm O/H} \sim
8.2 - 8.4$ \citep{Wolfire2010,Glover2010,Leroy2011}.  For our
range of metallicities ($\sim$$8.4 - 9.0$) neither the estimates of
\xco\ nor the metallicity measurements are accurate  enough that
we feel comfortable applying a correction to our data. Instead,
under the assumption of a fixed SFR--to--H$_2$ ratio, the observed
SFR--to--CO ratio can be utilized to constrain the metallicity dependence
of the \xco\ factor. \citet{Krumholz2011} adopt a version of this approach
using literature data and show that the observed metallicity variation
of the SFR--to--CO ratio is broadly consistent with \xco\ predicted by
\citet{Wolfire2010}, though with large scatter. We therefore expect
that \xco\ does affect our results, creating much of the observed
offset to higher SFR for low--mass galaxies. Subsequent analysis,
especially comparison to {\em Herschel} dust maps, will reveal if
there are also important second--order effects at play within galaxies.

Variations in the CO\jtwo/CO\jone\ line ratio create a second potential
bias in $\Sigma_{\rm H2}$. We adopt a fixed ratio of $0.7$ based on
comparison to literature CO\jone\ data. This is somewhat higher than
the observed ratio in the inner part of the Milky Way \citep{Fixsen1999},
$\sim$$0.5$, though the uncertainties on that ratio are large. More
important, \citet{Fixsen1999} suggest variations in the CO line ratios
between the inner and outer Milky Way and there are well--established
differences between normal disk and starburst galaxies. Although not
immediately apparent from a comparison of HERACLES to literature
CO\jone\ data \citep{Leroy2009a}, such variations could affect our
derived $\Sigma_{\rm H2}$ by as much as $\sim$50\%. Rosolowsky
et al. (2011, in prep.) will present a thorough investigation of how the
line ratio varies with environment in HERACLES.

\subsection{Distribution of Molecular Gas}
\label{sec:htwo}

% moved figure up in the text be achieve a better figure & text flow
% first reference appears in 3rd paragraph of Section 4.2.1 below
\begin{figure*}
\epsscale{1.0} \plottwo{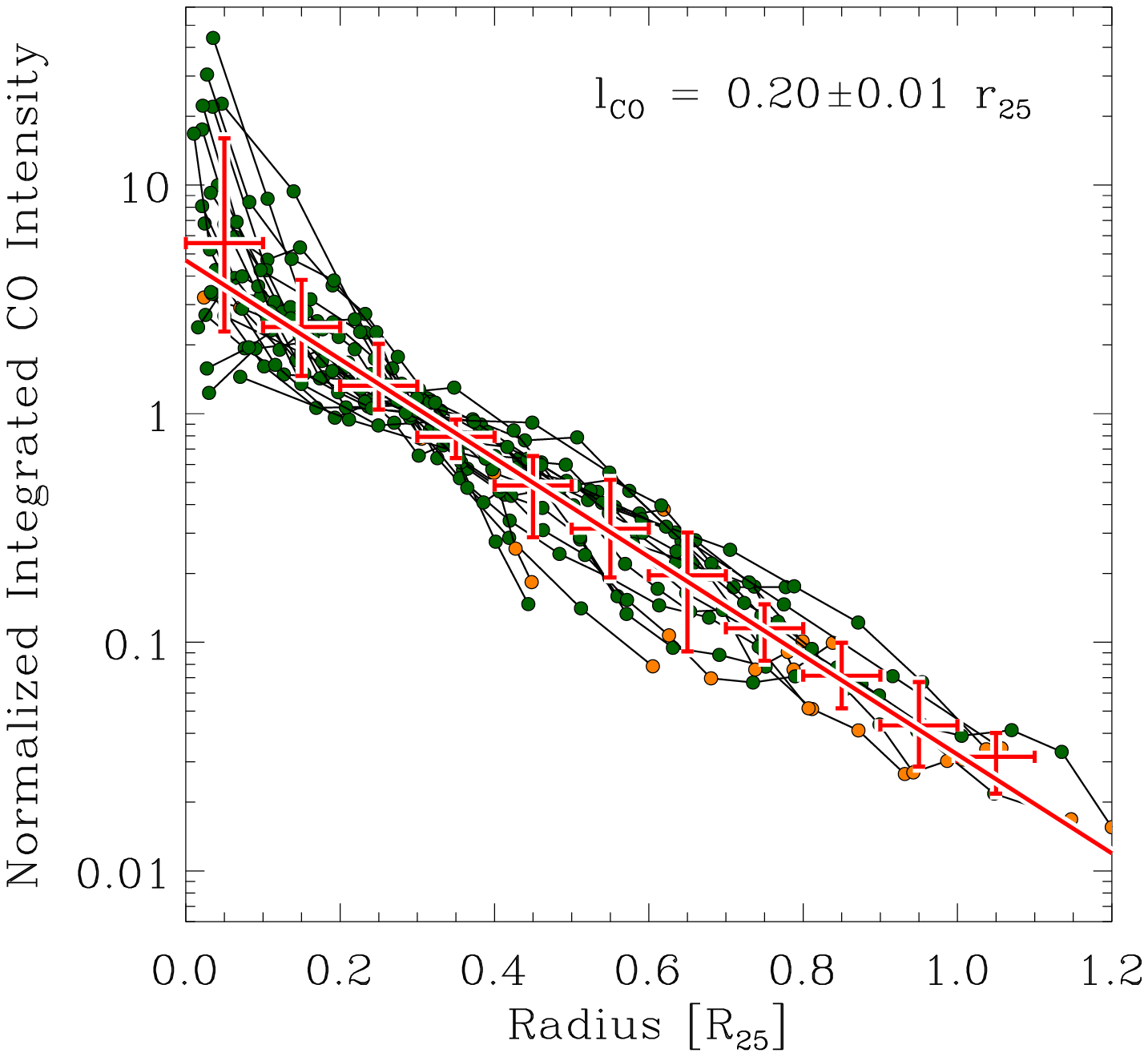}{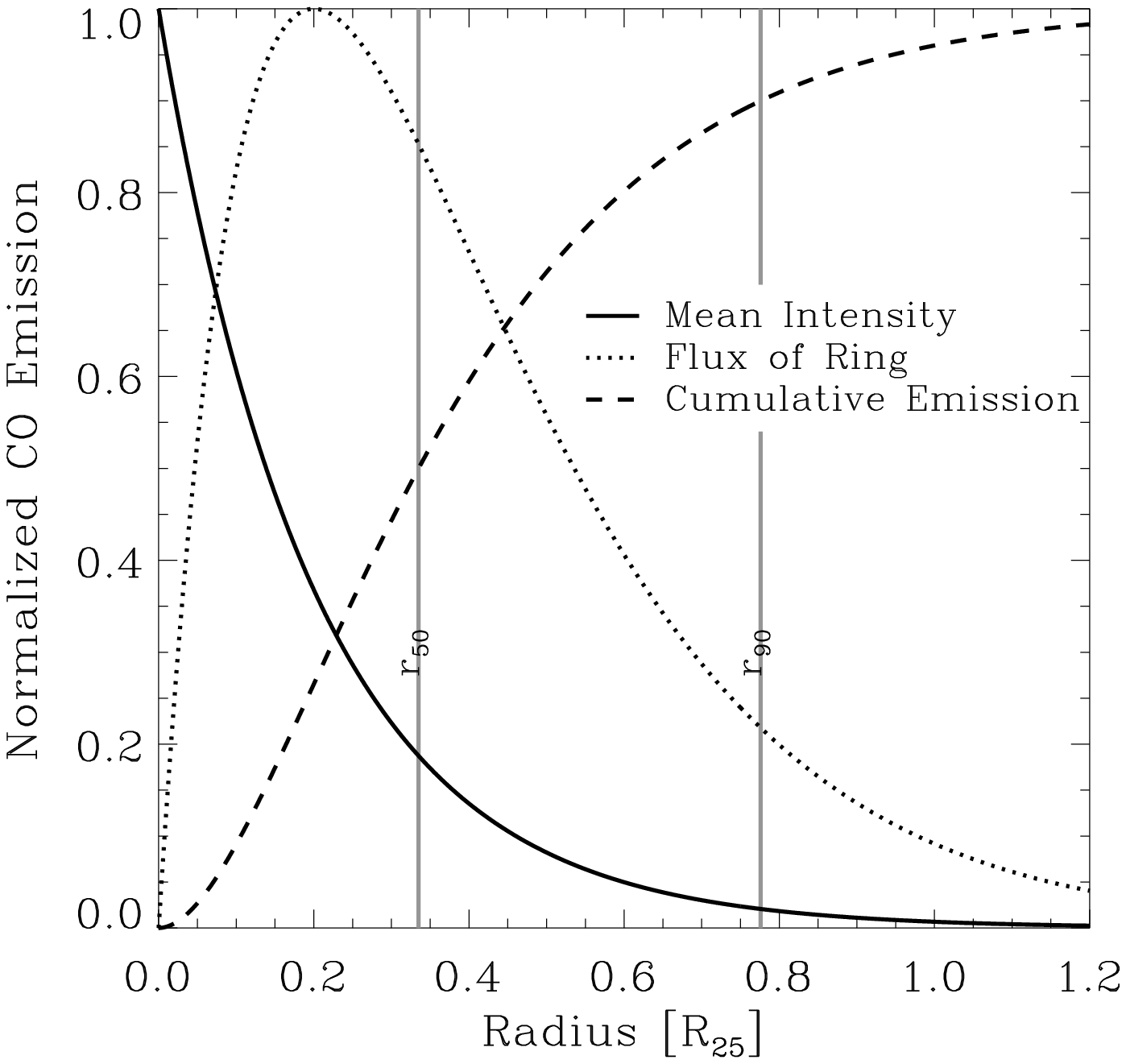}
\caption{({\em left}) Normalized CO intensity ($y$--axis) as a function
  of galactocentric radius ($x$--axis) in units of $r_{25}$, the 25$^{th}$
  magnitude $B$--band isophote for a respective galaxy. The CO
  intensity is normalized so that an exponential fit to the profile has
  intensity $1$ at $0.3~r_{25}$. Crosses mark the median and the
  $1\sigma$ scatter among galaxies in a series of radial bins (including
  only galaxies with at least marginal data in that bin). The average
  decline can be parametrized by an exponential with scale length
  of $0.2~r_{25}$ with no clear evidence of a truncation or break.
  ({\em right}) Schematic diagram for an exponential disk with scale
  length $0.2~r_{25}$. We plot mean intensity, flux in each ring, and
  enclosed luminosity as functions of radius. Gray lines show $r_{50}$
  and $r_{90}$, the radii at which 50 and 90\% of the flux are enclosed.
  \label{f12}}
\end{figure*}

The tight correlation between SFR tracers and CO emission across all
regimes strongly reinforces the primary importance of molecular gas to
star formation. In this section, we therefore examine the distribution of
molecular gas in galaxies. In the outer parts of spiral galaxies where
$\Sigma_{\rm HI} > \Sigma_{\rm H2}$, the formation of molecular gas
from atomic gas appears to represent the bottleneck to star formation
and the relative abundance of \htwo\ and \hi\ is key to setting the star
formation rate \citep[e.g.,][]{Leroy2008, Bigiel2010a}. We can apply
the large dynamic range in \htwo--to--\hi\ ratios achieved by stacking
to make improved measurements of how this key quantity varies across
galaxies.

\subsubsection{Radial Distribution of CO Intensity}
\label{sec:rad_prof}

Many previous studies have shown that azimuthally averaged CO
emission decreases with increasing galactocentric radius \citep[e.g.,][]
{Young1991, Young1995, Regan2001, Schuster2007}. Whereas
galaxy centers often exhibit deviations from the large scale trend, CO
emission outside the centers declines approximately uniformly with
radius \citep{Young1995}. A first analysis of the HERACLES data
revealed a characteristic exponential decline of CO emission in the
inner parts of galaxy disks, with the scale length of CO emission
similar to that of old stars and tracers of recent star formation
\citep{Leroy2009a}. With the increased sensitivity from stacking and
a larger sample, we can revisit this question and ask if this radial
decline in CO intensity continues smoothly out to $\sim$1 $r_{\rm 25}$.

From exponential fits to the high--significance CO data of each
galaxy\footnote{We exclude the following galaxies from the analysis
  because (a) their emission is compact compared to our beam:
  NGC 337, 3049, 3077, 4625; (b) they are barely detected: NGC
  4214, 4559; (c) their morphology is not well parametrized by an
  exponential: NGC 2798, 2976, 4725.} (solid--dashed lines in
Figure~\ref{f4} and Appendix), excluding the galaxy centers (inner
30\arcsec) we find a median exponential scale length, $l_{\rm CO}$,
of $0.21$ $r_{\rm 25}$ with $68$\% of all $l_{\rm CO}$ between
$0.16 - 0.28$ $r_{\rm 25}$. This value agrees well with typical
scale lengths found in previous studies \citep[e.g.,][]{Young1995}
and the individual galaxy scale lengths agree well with previous
work on the HERACLES sample \citep{Leroy2009a}.

The normalizations of these fits reflect galaxy--to--galaxy variations
in the total molecular gas content. In the left panel of Figure~\ref{f12}
we show all profiles aligned to a common normalization. We plot the
radius in units of $r_{\rm 25}$, the 25$^{th}$ magnitude $B$--band
isophote, and normalize each profile so that the exponential fits have
intensity $1$ at $r_{\rm gal} = 0.3$ $r_{\rm 25}$. The figure thus shows
the radial variation of CO intensity across our sample, controlled for
the overall CO luminosity and absolute size of each galaxy. Thick
crosses mark the median CO intensity and the $68$th percentile range
in bins $0.1$~$r_{\rm25}$ wide. The same exponential decline seen
in individual profiles is even more evident here, with $l_{\rm CO} =
0.20 \pm 0.01$ $r_{\rm25}$ for the average of the sample.

The left panel in Figure~\ref{f12} shows that the radial decline of the
CO profiles observed previously for the inner part of galaxies extends
without significant changes out to our last measured data points. In
most galaxies there is no clear evidence for a sharp cutoff or a change
in slope. As long as the normalized profiles are above the sensitivity
limit the decline appears to continue (without significant deviation) with
$l_{\rm CO} = 0.20$~$r_{\rm 25}$ on average. This smooth exponential
decline in CO intensity with increasing radius suggests that the observed
decline in star formation rate from the inner to outer parts of galaxy disks
is driven by a continuous decrease in the supply of molecular gas, rather
than a sharp threshold of some kind.

% moved figure up in the text be achieve a better figure & text flow
% first reference appears in 1st paragraph of Section 4.2.2 below
\begin{figure*}
\epsscale{1.0}
\plottwo{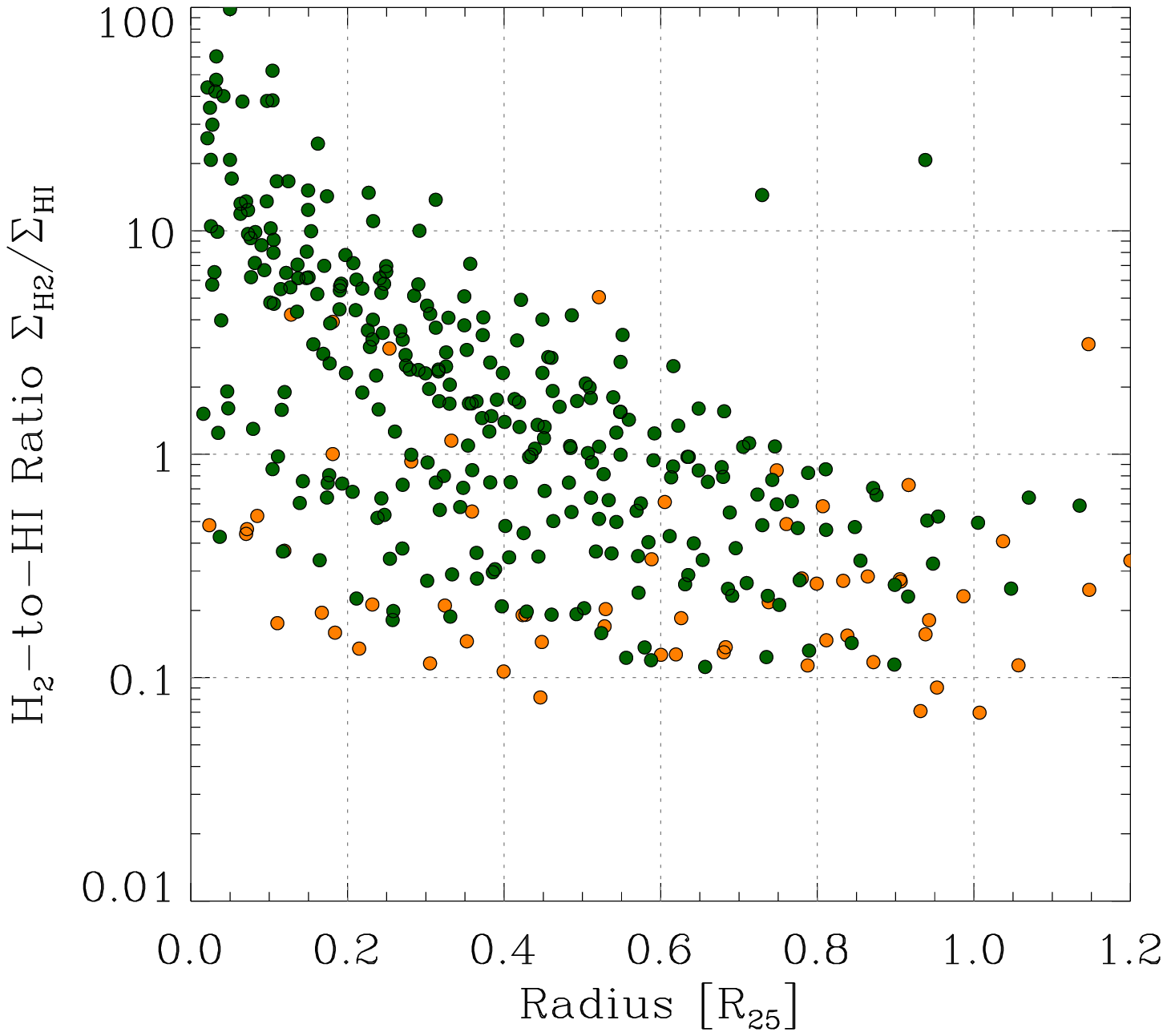}{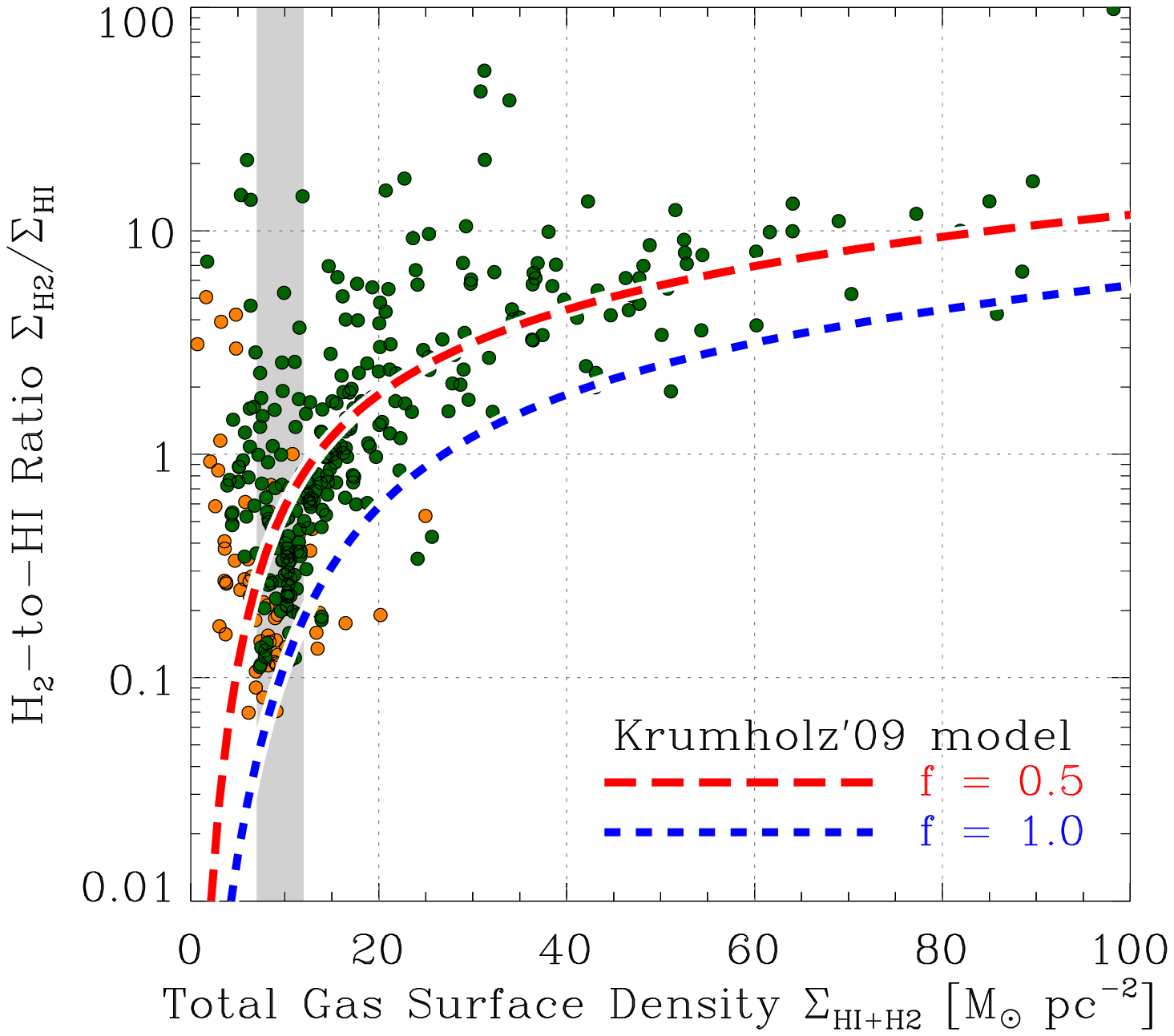}
\caption{The molecular--atomic ratio, $R_{\rm H2} = \Sigma_{\rm H2} /
  \Sigma_{\rm HI}$ as function of galactocentric radius (left panel), and
  total gas surface density, $\Sigma_{\rm HI+H2}$ (right panel). The
  dotted line in the right panel shows $R_{\rm H2}$ as predicted by a
  theoretical model by \citet[][see text]{Krumholz2009}. The gray shaded
  area indicates data that are shown as function of radius in Figure~\ref{f14}.
  We observe large variations in $R_{\rm H2}$ as function of radius and
  total gas surface density, however neither quantity is sufficient to
  parametrize the observed trend. \label{f13}}
\end{figure*}

There are several galaxies which deviate from this median exponential
trend. We already noted above that we do not fit exponential profiles to
NGC 2798, 2976, 4725 because their gas (\hi\ and CO) radial profiles
are insufficiently parametrized by exponentials. There are three galaxies
(NGC 2146, 2903, 4569) that we do fit and determine small exponential
scale lengths, $l_{\rm CO} \lesssim 0.1$~$r_{\rm 25}$. NGC 2146 hosts
an ongoing starburst and both NGC 2903 and NGC 4569 have prominent
bars that may be funneling molecular gas to their centers. For two other
galaxies (NGC 2841 and 4579) we determine large exponential scale
lengths, $l_{\rm CO} \gtrsim 0.3$~$r_{\rm 25}$. These galaxies are better
described by a flat distribution (or even a central depression) and a cutoff
at larger radii.

The tight correspondence of the CO scale length, $l_{\rm CO}$ to
$r_{25}$ has been noted before \citep{Young1995}, while other
studies have found a close correspondence between CO and
near--infrared light \citep{Regan2001,Blitz2004}. In our sample
there is a fairly good correspondence between $r_{25}$ and the
near--infrared scale length, $l_{\rm 3.6}$, measured at $3.6$~\micron\
with $r_{\rm 25} \approx 4.7 \pm 0.8~l_{\rm 3.6}$ \citep[][a treatment
of a larger sample suggests a slightly lower ratio of $\sim$$4.1$
with similar error bars]{Leroy2008}. The near--infrared light should
approximately trace the distribution of stellar mass, so that our
measured scale length is very similar to that of the stellar mass.
This tight coupling has been interpreted to indicate the importance
of the stellar potential well to collecting star--forming material
\citep{Blitz2006}. Here we see this correspondence to continue
into the regime where the molecular gas is not the dominant gas
component, confirming that molecular gas formation is the bottleneck
to star formation.

The right panel in Figure~\ref{f12} shows the distribution of
enclosed luminosity, mean intensity, and flux at each radius for an
exponential disk with scale length of $0.2~r_{25}$. The brightest
individual ring for such a disk lies at $0.2~r_{25}$ and half the flux
is enclosed within $r_{50} \approx 0.3~r_{25}$. This value, $r_{50}$,
is fairly close to the radius at which $\Sigma_{\rm HI} \approx 
\Sigma_{\rm H2}$ in a typical disk galaxy \citep[see the left panel
of Figure~\ref{f13} and][]{Leroy2008}, so that CO emission is about
evenly split between the H$_2$--dominated and \hi --dominated
parts of such a galaxy. Meanwhile, 90\% of the flux lies within $r_{90}
\sim 0.8~r_{25}$ a value that is very similar to the threshold radius
identified by \citet{Martin2001}. We do not find evidence to support
a true break at this radius, but as an ``edge'' to the star--forming disk,
a 90\% contour may have utility.

\subsubsection{The H$_2$--to--HI Ratio}
\label{sec:rh2}

In the outer parts of galaxy disks --- and thus over most of the area
in galaxy disks --- we have $\Sigma_{\rm HI} \gtrsim \Sigma_{\rm H2}$,
implying that star--forming H$_2$ gas does not make up most of
the interstellar medium. In this regime the relative abundance of
H$_2$ and \hi\ is a key quantity to regulate the star formation rate.
Observations over the last decade have revealed strong variations of
the fraction of gas in the molecular phase as a function of
galactocentric radius, total gas surface density, stellar surface
density, disk orbital time, and interstellar pressure \citep{Wong2002,
  Heyer2004, Blitz2006, Leroy2008, Wong2009}. In Figure~\ref{f13}
we show the two most basic of these trends, the H$_2$--to--\hi\ ratio,
$R_{\rm H2} = \Sigma_{\rm H2}/\Sigma_{\rm HI}$, as a function of
normalized galactocentric radius (left panel) and total gas surface
density (right panel). We focus on $R_{\rm H2}$ because it is more
easily separated in discrete observables than the fraction of gas that
is molecular, $f_{\rm H2} = \Sigma_{\rm H2} / (\Sigma_{\rm HI} +
\Sigma_{\rm H2}) = R_{\rm H2} / (1 + R_{\rm H2})$.  This makes it
easier to interpret uncertainties and systematic effects like changes
in the CO--to--H$_2$ conversion factor.

The left panel of Figure~\ref{f13} shows $R_{\rm H2}$ as function of
galactocentric radius in units of $r_{25}$ for all data detected with
high or marginal significance. For clarity, we do not plot $R_{\rm H2}$
for regions where we determined only upper limits in CO intensity;
for the inner parts ($r < 0.6~r_{25}$) these upper limits are bounded
by $R_{\rm H2} \lesssim 0.3$, whereas for outer parts ($r > 0.6~r_{25}$)
the upper limits in $\Sigma_{\rm H2}$ are typically of comparable
magnitude as measurements of $\Sigma_{\rm HI}$ and upper limits
are bounded by $R_{\rm H2} \lesssim 1$. In agreement with
\citet{Wong2002}, \citet{Heyer2004}, \citet{Bigiel2008}, and
\citet{Leroy2008} we find $R_{\rm H2}$ to decline with increasing
galactocentric radius, a variation that reflects the distinct radial
profiles of atomic and molecular gas. However, radial variations
alone do not explain the full range of observed molecular fractions
because $R_{\rm H2}$ can vary by up to two orders of magnitude
at any given galactocentric radius.

A significant part of the variations in $R_{\rm H2}$ at a given
galactocentric radius corresponds to systematic variations between
galaxies. These are mainly caused by variations in the absolute
molecular gas content of a galaxy and can be removed by a
normalization procedure similar to one applied in the left panel
of  Figure~\ref{f12}. The result is similar to that seen for the
radial profiles of CO: the relationship tightens and we can see that
most of the decline of $R_{\rm H2}$ inside a galaxy occurs radially.
However, there is significantly more scatter remaining in $R_{\rm H2}$
versus radius than we observed for the normalized CO radial profiles,
highlighting the importance of parameters other than a combination
of radius and host galaxy to set $R_{\rm H2}$.

In addition to declining with increasing galactocentric radius,
$R_{\rm H2}$ increases as the total gas surface density, $\Sigma_{\rm
  gas} = \Sigma_{\rm HI} + \Sigma_{\rm H2}$, increases. The more gas
that is present along a line of sight, the larger the fraction of gas that is
molecular. The right panel of Figure~\ref{f13} shows this result,
plotting $R_{\rm H2}$ as function of the total gas surface density,
$\Sigma_{\rm gas}$. $R_{\rm H2}$ increases with increasing
$\Sigma_{\rm gas}$, with regions of high surface density,
$\Sigma_{\rm gas} \gtrsim 20$ \Msunperpc, being predominately
molecular, $R_{\rm H2} \gtrsim 1$. 

At high $\Sigma_{\rm gas}$, the right panel of Figure~\ref{f13} is
largely a way of visualizing the ``saturation'' of $\Sigma_{\rm HI}$
on large scales in galaxies. A number of authors have found that
averaged over hundreds of parsecs to kpc scales, $\Sigma_{\rm
HI}$ rarely exceeds $\sim$10 \Msunperpc\ \citep{Wong2002}.
This limit is clearly violated at high spatial resolution \citep[e.g.][]
{Kim1999, Stanimirovic1999, Brinks1984} and may vary among
classes of galaxies. Several recent theoretical works have aimed
at reproducing this behavior. \citet{Krumholz2009} focused on
shielding of H$_2$ inside individual atomic--molecular complexes,
whereas \citet{Ostriker2010} examined the interplay between
large--scale thermal and dynamical equilibria.

The right panel in Figure~\ref{f13} includes also the predicted
solar--metallicity $\Sigma_{\rm gas}$--$R_{\rm H2}$ relation
from \citet[][their Equations 38 \& 39]{Krumholz2009}. They
model individual atomic--molecular complexes, however these
complexes have a filling factor substantially less than $1$ inside
our beam. The appropriate $\Sigma_{\rm gas}$ to input into their
model is therefore the average surface densities of the complexes,
$\Sigma_{\rm comp}$, within our beam, which will be related to our
observed surface density, $\Sigma_{\rm gas}$, by the filling factor,
$f$, namely: $\Sigma_{\rm gas} = f \ \Sigma_{\rm comp}$. The
model curve (blue short--dashed line) with a filling factor $f=1$ is
offset towards higher $\Sigma_{\rm gas}$ (shifted right) or lower
$R_{\rm H2}$ (shifted down) compared to our data. To have the
model curve intersect our data (red long--dashed line) requires
a filling factor of $f \approx 0.5 - 1$, so that atomic--molecular
complexes fill about half the areas in our beam. This is before
any accounting for the presence of diffuse \hi\ not in star--forming
atomic--molecular complexes and is assuming a fixed filling factor,
both of which are likely oversimplifications. Nonetheless, the
\citet{Krumholz2009} curve does show an overall good
correspondence to our data.

\begin{figure}
\epsscale{1.0}
\plotone{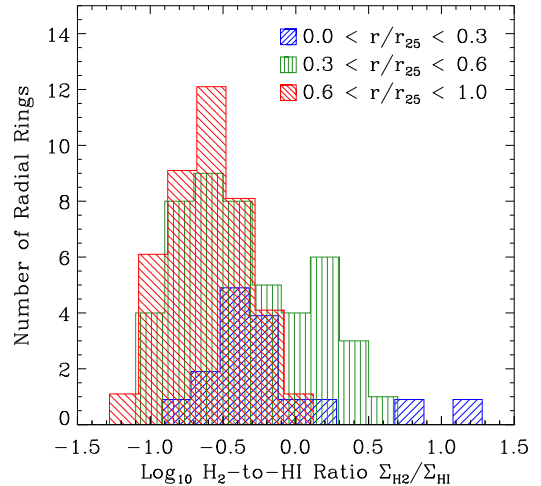}
\caption{The radial dependence of $R_{\rm H2}$ for regions with
  comparable total gas surface density, $\Sigma_{\rm HI+H2} =  7-12$
  \Msunperpc . The data are taken from the gray highlighted region in the
  left panel of Figure~\ref{f13}. $R_{\rm H2}$ is not sufficiently determined
  by the total gas surface density but shows also a radial dependence.
  \label{f14}}
\end{figure}

As with the radius, the total gas surface density predicts some of the
broad behavior of $R_{\rm H2}$ but knowing $\Sigma_{\rm gas}$
does not uniquely specify the amount of molecular gas, particularly
at low $\Sigma_{\rm gas}$. This are visible as the large scatter in
$R_{\rm H2}$ at low surface densities in the right panel of
Figure~\ref{f13}. The scatter reflects a dependence of $R_{\rm H2}$
on environmental factors other than gas surface density.
Figure~\ref{f14} shows $R_{\rm H2}$ over the small range
$\Sigma_{\rm gas} = 7 - 12$ \Msunperpc , i.e., the data from the gray
highlighted region in Figure~\ref{f13}. We plot histograms for several
radial bins which are clearly offset, indicating an additional radial
dependence of $R_{\rm H2}$. At small radii ($r \lesssim 0.3$
$r_{\rm 25}$) we observe a large scatter in $R_{\rm H2}$ for
$\Sigma_{\rm gas} = 7 - 12$~\Msunperpc\ whereas at large radii
($r \gtrsim 0.6$ $r_{\rm 25}$) gas with this surface density is
always predominantly \hi .

As was the case in SFR--H$_2$ space, distinct populations of galaxies are
responsible for some of the variations in the right panel of Figure~\ref{f13}.
Early type (Sab--Sb) spirals (e.g., NGC 2841, 3351, 3627, 4736) often show
large molecular fractions, $R_{\rm H2} \gtrsim 0.5$, but typically have low
\hi\ and \htwo\ surface densities, $\Sigma_{\rm HI}$ and $\Sigma_{\rm H2}
\approx 1 - 5$ \Msunperpc . By contrast, massive Sc galaxies (e.g., NGC
4254, 4321, 5194, 6946) can have comparably high molecular fractions,
$R_{\rm H2} \approx 0.35 - 1$, but have higher surface density \hi\ disks,
$\Sigma_{\rm HI} \approx 5 - 10$ \Msunperpc , so that these fractions occur
at higher $\Sigma_{\rm gas}$. A trend with metallicity is not immediately
obvious in the data but these differences may reflect the more substantial
stellar surface densities found in the earlier--type galaxies. This increased
stellar surface density results in a stronger gravitational field, which could
lead to a higher midplane gas pressure \citep{Blitz2006} and a low fraction
of diffuse \hi\ gas \citep{Ostriker2010}.

\subsubsection{Discussion of $R_{\rm H2}$}

Following several recent studies we observe strong systematic
variations in the H$_2$--\hi\ balance across galaxies. Two of the
strongest behaviors are an approximately exponential decrease
in $R_{\rm H2}$ with increasing galactocentric radius and a steady
increase in $R_{\rm H2}$ with increasing gas surface density. Our
improved sensitivity shows these trends extending to low surface
densities and our expanded sample makes clear that neither of
these basic parametrizations adequately captures the entire range
of $R_{\rm H2}$ variations. The likely physical drivers for the scatter
in $R_{\rm H2}$ that we observe are metallicity and dust--to--gas
ratio \citep[e.g.,][Bolatto et al. 2011, in prep.]{Leroy2008, Krumholz2009,
  Gnedin2009}, the dissociating radiation field \citep[e.g.,][]
  {Robertson2008}, variations in interstellar gas pressure and
density \citep[e.g.,][]{Elmegreen1994, Wong2002, Blitz2006}, and
external perturbations that drive gas to higher densities \citep[e.g.,][]
  {Blitz2006, Bournaud2010}. Each of these quantities are
observationally accessible in our sample \citep[e.g.,][]{Leroy2008}
and estimates of the relative roles of each process will be presented
in Leroy et al. (2011, in prep.).

\begin{figure}
\epsscale{1.0} \plotone{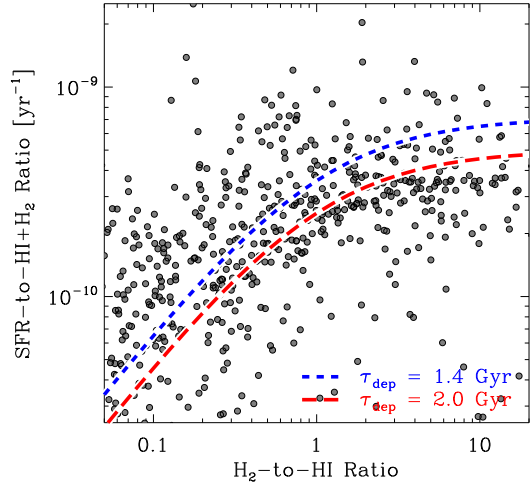}
\caption{The recent star formation rate per unit total gas, $\Sigma_{\rm SFR}
  /  \Sigma_{\rm HI+H2}$ (the inverse of the total gas depletion time;
  \mbox{$y$--axis}), as function of the molecular--atomic gas ratio,
  $R_{\rm H2} = \Sigma_{\rm H2} / \Sigma_{\rm HI}$ \mbox{($x$--axis)}.
  A fixed ratio of  SFR--to--H$_2$ is shown by the blue dotted line for a
  {\em molecular} gas depletion time of $\tau_{\rm dep} \approx 1.4$~Gyr
  and by the red dashed line for $\tau_{\rm dep} \approx 2$~Gyr. In regions
  dominated by molecular gas ($R_{\rm H2} > 1$) the $\Sigma_{\rm SFR}
  /  \Sigma_{\rm HI+H2}$ ratio approximates a value corresponding to the
  constant molecular depletion time, $\tau_{\rm dep}$. In regions of small
  molecular gas fraction ($R_{\rm H2} < 1$) the $\Sigma_{\rm SFR} / 
  \Sigma_{\rm HI+H2}$ ratio decreases significantly indicating that the
  total gas does not scale uniformly with the recent star formation rate. 
  \label{f15} }
\end{figure}

The overall relationship of total gas and star formation rate (middle
panels of Figure~\ref{f11}) can be reproduced by a roughly fixed
ratio of SFR--to--H$_2$ within galaxies (Section~\ref{sec:gastype})
and the observed scaling of $R_{\rm H2}$ with $\Sigma_{\rm HI+H2}$
(Section~\ref{sec:rh2}).  This is shown in Figure~\ref{f15}, where we
plot the SFR per unit total gas, $\Sigma_{\rm SFR} /  \Sigma_{\rm HI+H2}$
(the inverse of the {\em total} gas depletion time) as function of the
molecular--to--atomic ratio, $R_{\rm H2} = \Sigma_{\rm H2} / \Sigma_{\rm HI}$.
At low surface densities, the $R_{\rm H2}$--$\Sigma_{\rm HI+H2}$ relation
regulates $\Sigma_{\rm SFR}$, at high surface densities where almost
all of the gas is molecular the SFR--H$_2$ scaling determines the observed
ratio. This trend is well parametrized by a fixed SFR--to--H$_2$ ratio with
a {\em molecular} gas depletion time of $\tau_{\rm dep} \approx 1.4$~Gyr
(blue dotted line) for all of our targets and $\tau_{\rm dep} \approx 2$~Gyr
(red dashed line) for big spirals \citep[i.e. using the same sample as][]
{Bigiel2008,Leroy2008}. The transition between these two regimes creates
the curved shape seen in Figure~\ref{f11} \& \ref{f15}. This offers more
support for a modified version of the classical picture of a star formation
threshold \citep{Skillman1987, Kennicutt1989, Martin2001}, in which dense,
mostly molecular gas forms stars but the efficiency with which such (molecular)
gas forms is a strong function of environment, decreasing steadily with
decreasing gas surface density and increasing galactocentric radius.

\section{Summary}
\label{sec:summary}

We combine HERACLES CO\jtwo\ data with \hi\ velocity fields, mostly
from THINGS, to make sensitive measurements of CO intensity across
the disks of 33 nearby star--forming galaxies. We stack CO spectra across
many lines of sight by assuming that the mean \hi\ and CO velocities are
similar, an assumption that we verify in the inner parts of galaxies. This
approach allows us to detect CO out to galactocentric radii $\sim$1
$r_{\rm 25}$. Because we measure integrated CO intensities as low as
$0.3$ \Kkmpers\ ($\sim$1 \Msunperpc , before any correction for inclination)
with high significance, we are able to robustly measure CO intensities in
parts of galaxies where most of the ISM is atomic.

Using this approach we compare the radially averaged intensities of
FUV, \ha, IR, CO, and \hi\ emission across galaxy disks. We find an
approximately linear relation between CO intensity and monochromatic
IR intensity at both 24~\micron\ and 70~\micron . For the first time, we show
that these scaling relations continue smoothly from the H$_2$--dominated
to \hi--dominated ISM. Extinction causes FUV and \ha\ emission
to display a more complex relationship with CO, especially in the inner
parts of galaxies. In the outer parts of galaxy disks FUV and \ha\ emission
do correlate tightly with CO emission after galaxy--to--galaxy variations
are removed.

We use two calibration to estimate the recent star formation rate,
FUV$+$24\micron\ and \ha$+$24\micron , which we compare to H$_2$
derived from CO. We find an approximately linear relation between
$\Sigma_{\rm SFR}$ and $\Sigma_{\rm H2}$ in the range of
$\Sigma_{\rm H2} \approx 1 - 100$ \Msunperpc\ with no notable
variation between the two SFR estimates. A number of recent studies
\citep{Bigiel2008, Leroy2008, Blanc2009, Bigiel2011} have also seen
a roughly linear relationship between $\Sigma_{\rm SFR}$ and
$\Sigma_{\rm H2}$ and have argued that it implies that the surface
density of H$_2$ averaged over large scales does not strongly affect
the efficiency with which molecular gas forms stars.

We do find evidence for variations in the SFR--to--CO ratio among
galaxies. Indeed, most of the scatter in the relations between CO and
SFR tracers is driven by galaxy--to--galaxy variations. These variations
are not random, but show the trend observed by \citet{Young1996}
that lower mass, lower metallicity galaxies have higher ratios of
SFR--to--H$_2$ than massive disk galaxies. It will take further study
to determine whether these are real variations in the efficiency of star
formation or reflect changes in the CO--to--H$_2$ conversion factor
due to lower metallicities in these systems. After removing these
galaxy--to--galaxy variations the composite H$_2$--SFR relation is
remarkably tight, reinforcing a close link between H$_2$ and star
formation inside galaxies.

We compare the scaling between the surface densities of SFR and
\hi , \htwo , and total gas (\hi+\htwo). The relationship between SFR
and total gas has roughly the same rank correlation coefficient as
that between SFR and H$_2$, but does not obey a single functional
form. Where $\Sigma_{\rm HI} > \Sigma_{\rm H2}$ the relationship
between $\Sigma_{\rm SFR}$ and $\Sigma_{\rm HI+H2}$ is steep
whereas where $\Sigma_{\rm HI} < \Sigma_{\rm H2}$ the relationship
is much flatter. Meanwhile, we observe a linear relationship between
$\Sigma_{\rm H2}$ and $\Sigma_{\rm SFR}$ for the full range of
$\Sigma_{\rm H2} = 1 - 100$ \Msunperpc .  $\Sigma_{\rm HI}$ and
$\Sigma_{\rm SFR}$ are weakly correlated and exhibit a strongly
nonlinear relation, except at very large radii.

The unbroken extension of the $\Sigma_{\rm SFR}$--$\Sigma_{\rm H2}$
relation into the \hi --dominated regime suggests a modified version of the
classical picture of a star formation threshold \citep{Skillman1987,
Kennicutt1989, Martin2001}, in which stars form at fixed efficiency out of
molecular gas, to first order independent of environment within a galaxy.
The observed turn--over in the relation between SFR and {\em total} gas
relates to the H$_2$--to--\hi\ ratio which is a strong function of environment.

We therefore investigate the distribution of H$_2$ traced by CO using
our stacked data and compare it to the \hi . On large scales we observe
CO to decrease exponentially with a remarkably uniform scale length of
$\sim$$0.2~r_{\rm 25}$, again extending previous studies to lower
surface densities. We find the normalization of this exponential decline
to vary significantly among galaxies. The H$_2$--to--\hi\ ratio, traced
by the ratio of CO--to--\hi\ intensities, also varies systematically across
galaxies. It exhibits significant correlations with both galactocentric
radius and total gas surface density and we present high--sensitivity
measurements of both of these relationships. However, neither quantity
is sufficient to uniquely predict the H$_2$--to--\hi\ ratio on its own.

\acknowledgements

We thank the teams of SINGS, LVL, and GALEX NGS for making their
outstanding data sets available. We thank Daniela Calzetti for helpful
comments. We thank the anonymous referee for thoughtful comments
that improved the paper. The work of AS was supported by the Deutsche
Forschungsmeinschaft (DFG) Priority Program 1177. Support for AKL was
provided by NASA through Hubble Fellowship grant HST-HF-51258.01-A
awarded by the Space Telescope Science Institute, which is operated by
the Association of Universities for Research in Astronomy, Inc., for NASA,
under contract NAS 5-26555. The work of WJGbB is based upon research
supported by the South African Research Chairs Initiative of the Department
of Science and Technology and the National Research Foundation. This
research made use of the NASA/IPAC Extragalactic Database (NED),
which is operated by the JPL/Caltech, under contract with NASA, NASAÕs
Astrophysical Data System (ADS), and the HyperLeda catalog, located
on the WorldWide Web at http://www.obs.univ-lyon1.fr/hypercat/intro.html.

\begin{appendix}

Here we present the radial profiles of CO, \hi , FUV, \ha , and IR at
24~\micron\ and 70~\micron\ used to generate the plots in this paper.
We average the data in 15\arcsec\ wide tilted rings. For the CO data,
we stack the shifted spectra over this area and determine the integrated
CO intensity from fitting line profiles to the stacked spectrum. For the
\hi , FUV, \ha , and IR data we use two--dimensional maps of intensity
(Section~\ref{sec:method}). We determine the $1\sigma$ scatter from
the 68$^{\rm th}$--percentile from the data inside each ring. We plot
these as error bars but note the distinction from the uncertainty in the
mean.

For each galaxy we present two plots: The left panel shows CO and \hi\
both in units of observed intensities (\Kkmpers , left--hand $y$--axis) and
converted to mass surface densities (\Msunperpc , right--hand $y$--axis)
of H$_2$ and \hi . The color of the CO points indicates the significance
with which we could determine the integrated CO intensities: green for
high significance measurements, orange for measurements of marginal
significance and red for $3\sigma$ upper limits. To have H$_2$ and \hi\
on the same mass surface density scale, we multiplied the observed 21-cm
line intensities by a factor of $312.5$ (the ratio of Equations~\eqref{eq:hi}
and \eqref{eq:h2}). We also plot the star formation rate (SFR) surface
density (\Msunperyrperkpc ) determined from \ha$+$24\micron\ and
FUV$+$24\micron . Black solid-dashed lines show our exponential fit
to the radial CO profile. We fit all high significance data excluding galaxy
centers, defined as the the inner 30\arcsec , which often exhibit breaks
from the overall profile \citep{Regan2001,Helfer2003}. The derived
exponential scale lengths (in units of $r_{25}$, the radius of the 25$^{th}$
magnitude $B$--Band isophote), appear in the lower left corner.

The right panel shows observed intensities (in \mbox{MJy~sr$^{-1}$})
of our SFR tracers --- \ha , FUV, 24~\micron\ and 70~\micron\ emission.

\renewcommand{\thefigure}{\alph{figure}}
\setcounter{figure}{0}
\begin{figure*}
	\epsscale{1.0} \plotone{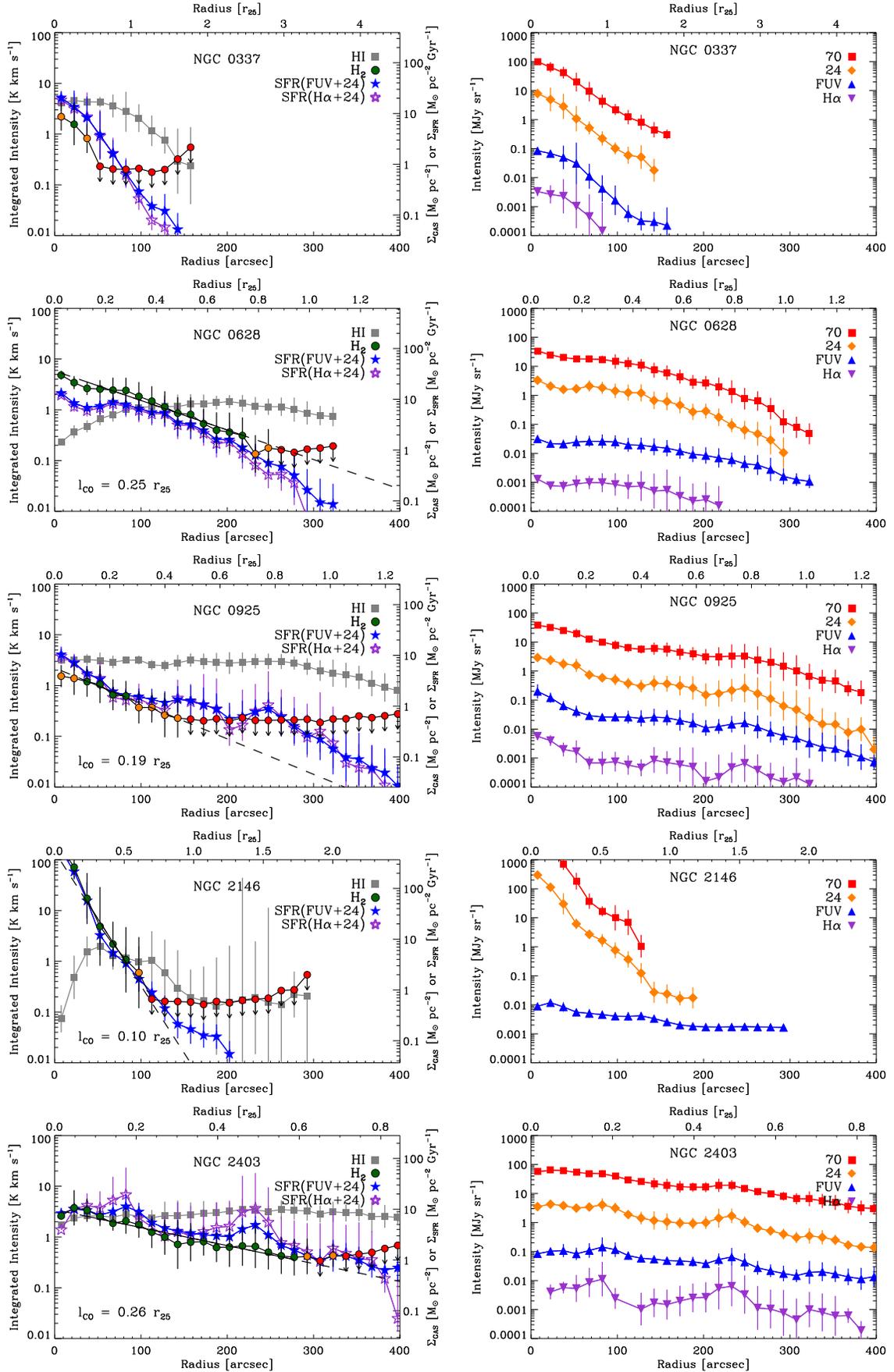}
	\caption{\label{f16a}Atlas of radial profiles, see Figure~\ref{f4} for details.}
\end{figure*}

\setcounter{figure}{0}
\begin{figure*}
	\epsscale{1.0} \plotone{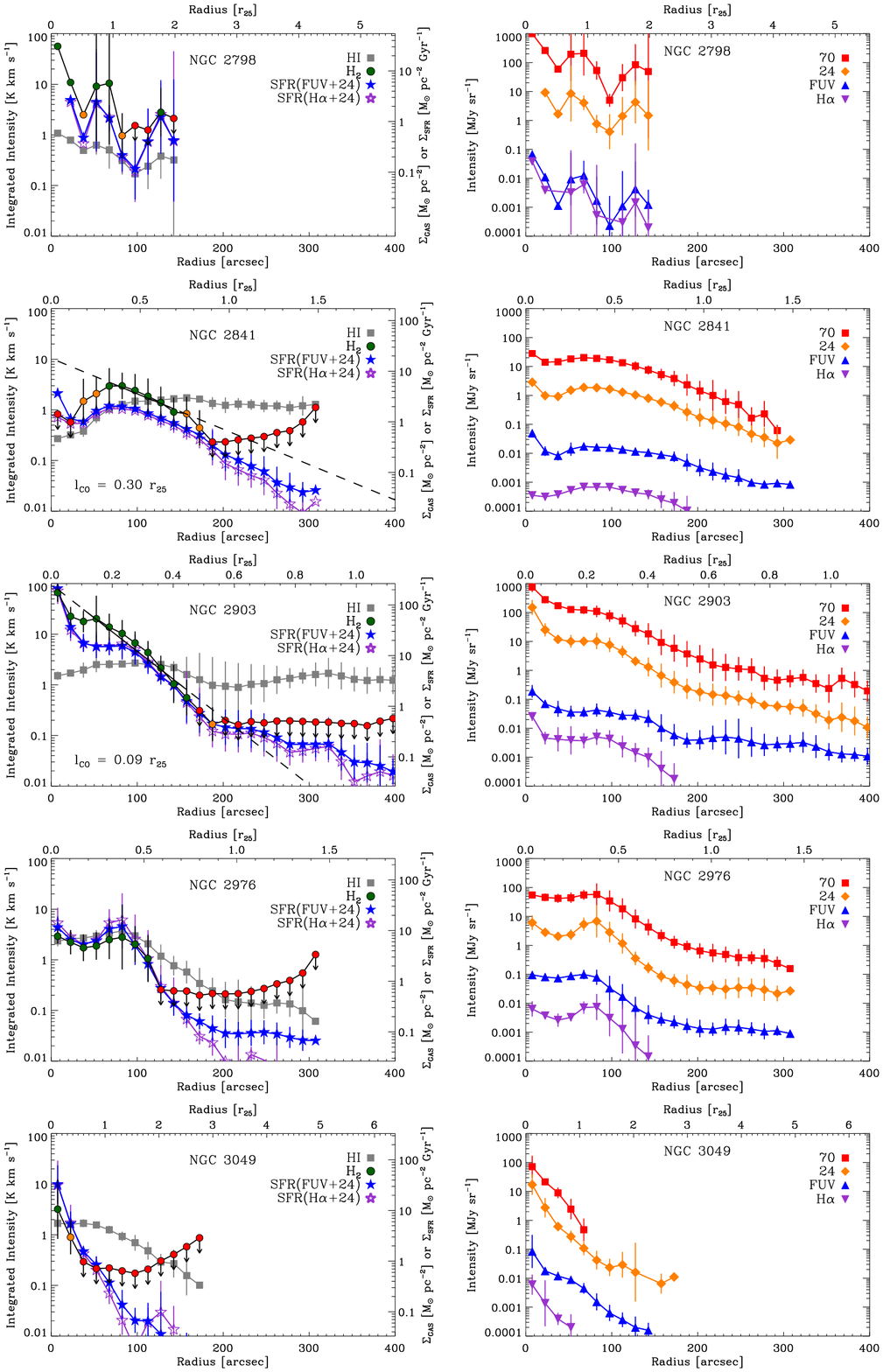}
	\caption{\label{f16b}Atlas of radial profiles, see Figure~\ref{f4} for details.}
\end{figure*}

\setcounter{figure}{0}
\begin{figure*}
	\epsscale{1.0} \plotone{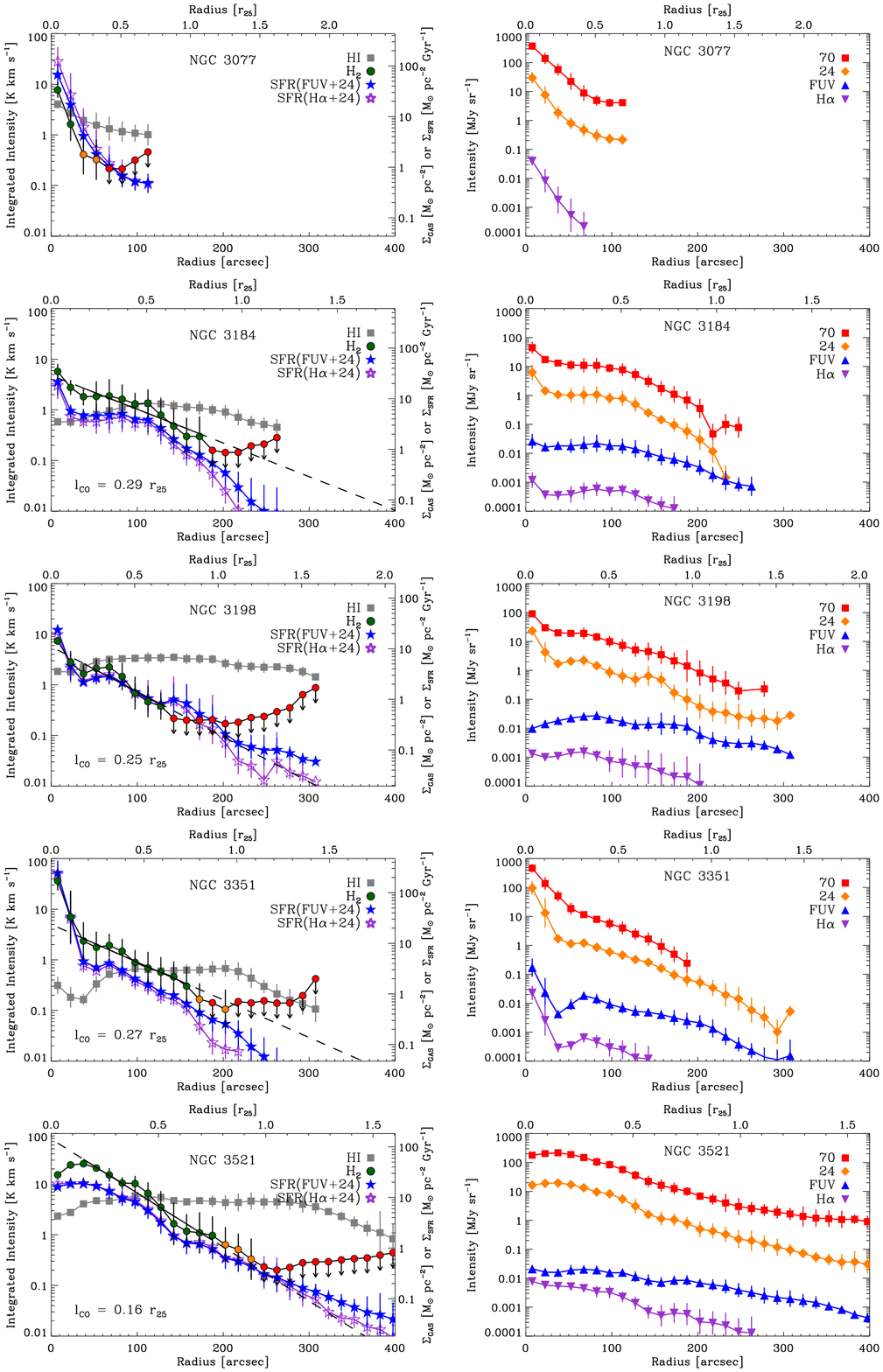}
	\caption{\label{f16c}Atlas of radial profiles, see Figure~\ref{f4} for details.}
\end{figure*}

\setcounter{figure}{0}
\begin{figure*}
	\epsscale{1.0} \plotone{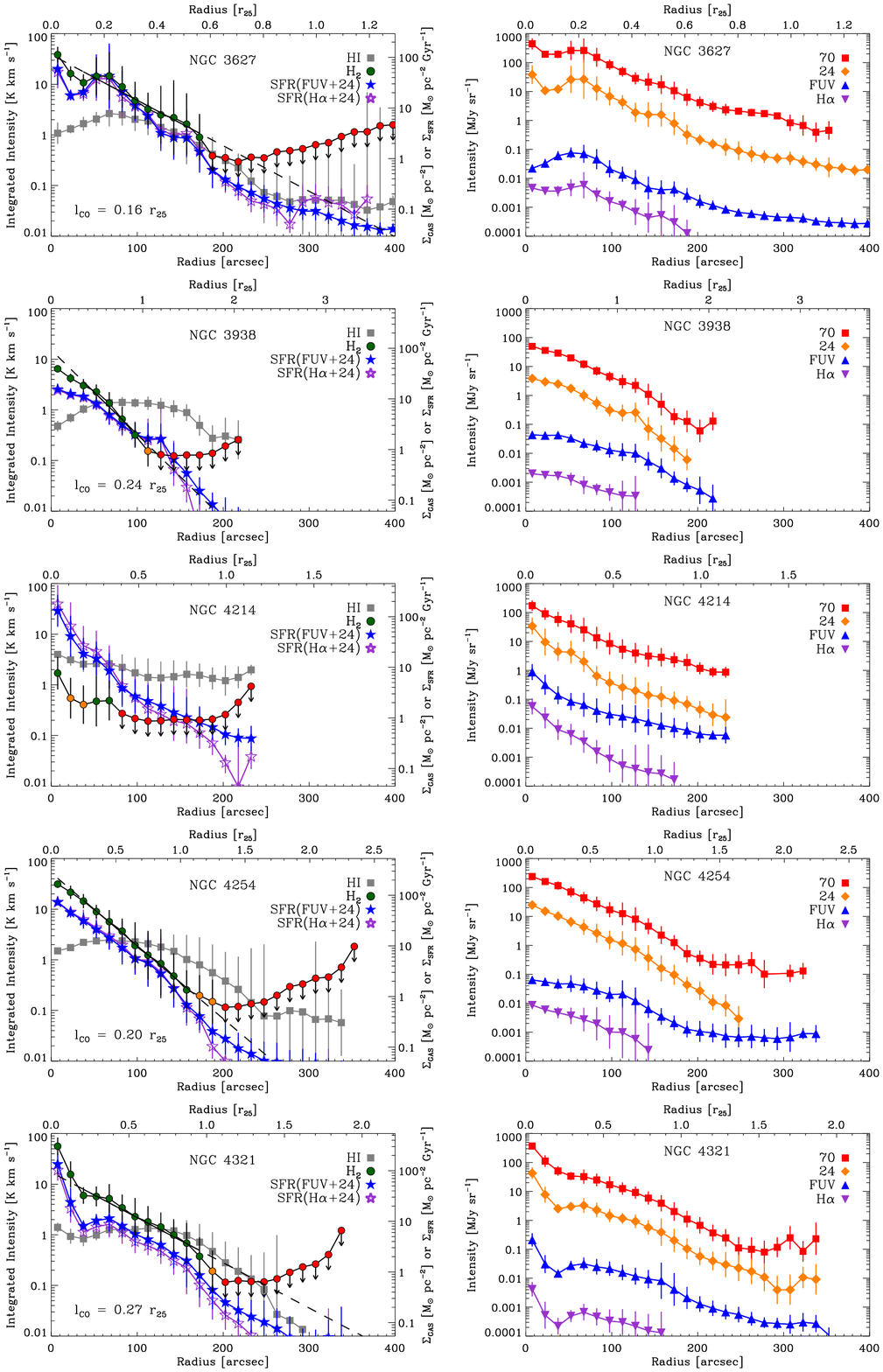}
	\caption{\label{f16d}Atlas of radial profiles, see Figure~\ref{f4} for details.}
\end{figure*}

\setcounter{figure}{0}
\begin{figure*}
	\epsscale{1.0} \plotone{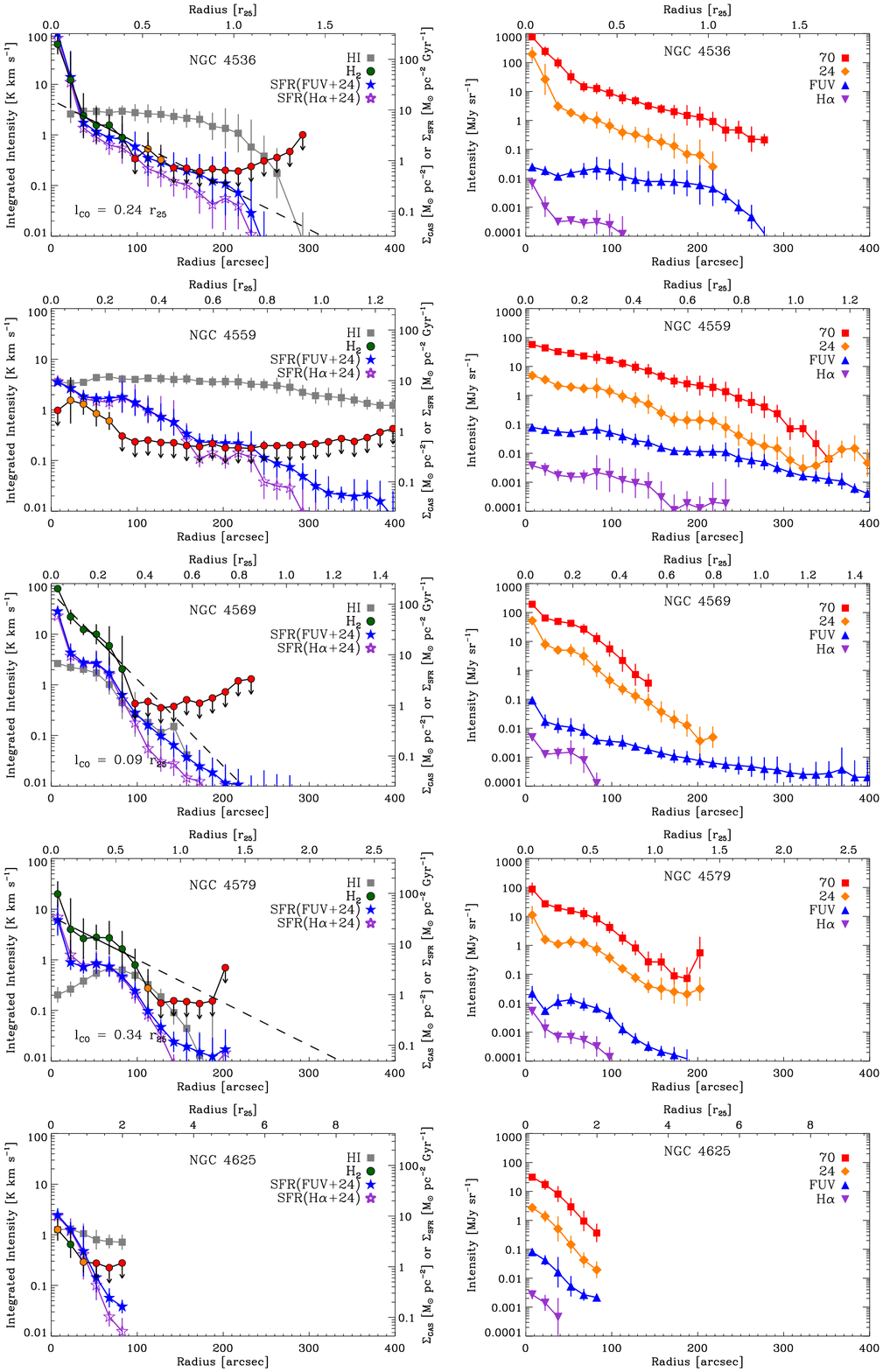}
	\caption{\label{f16e}Atlas of radial profiles, see Figure~\ref{f4} for details.}
\end{figure*}

\setcounter{figure}{0}
\begin{figure*}
	\epsscale{1.0} \plotone{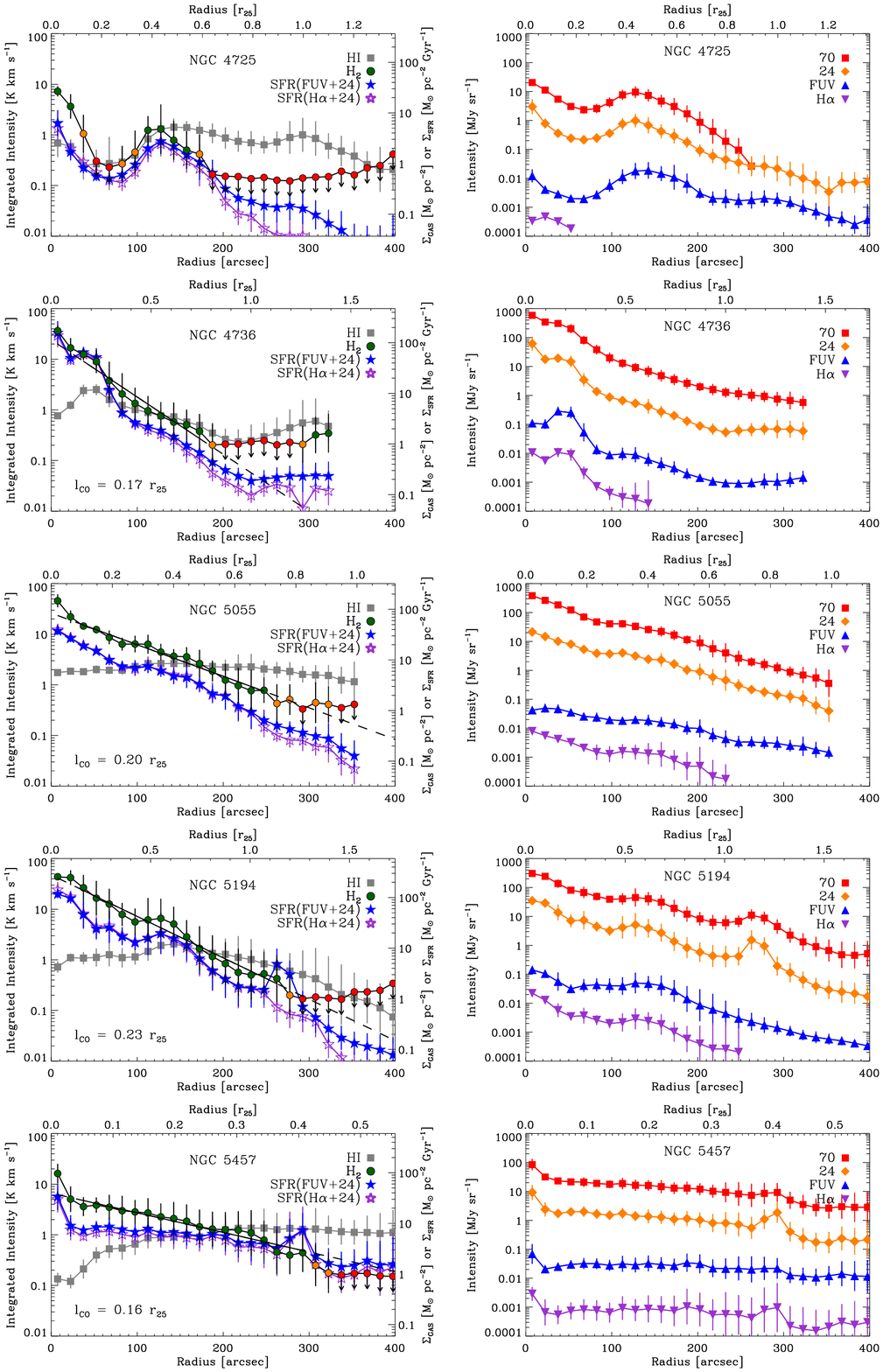}
	\caption{\label{f16f}Atlas of radial profiles, see Figure~\ref{f4} for details.}
\end{figure*}

\setcounter{figure}{0}
\begin{figure*}
	\epsscale{1.0} \plotone{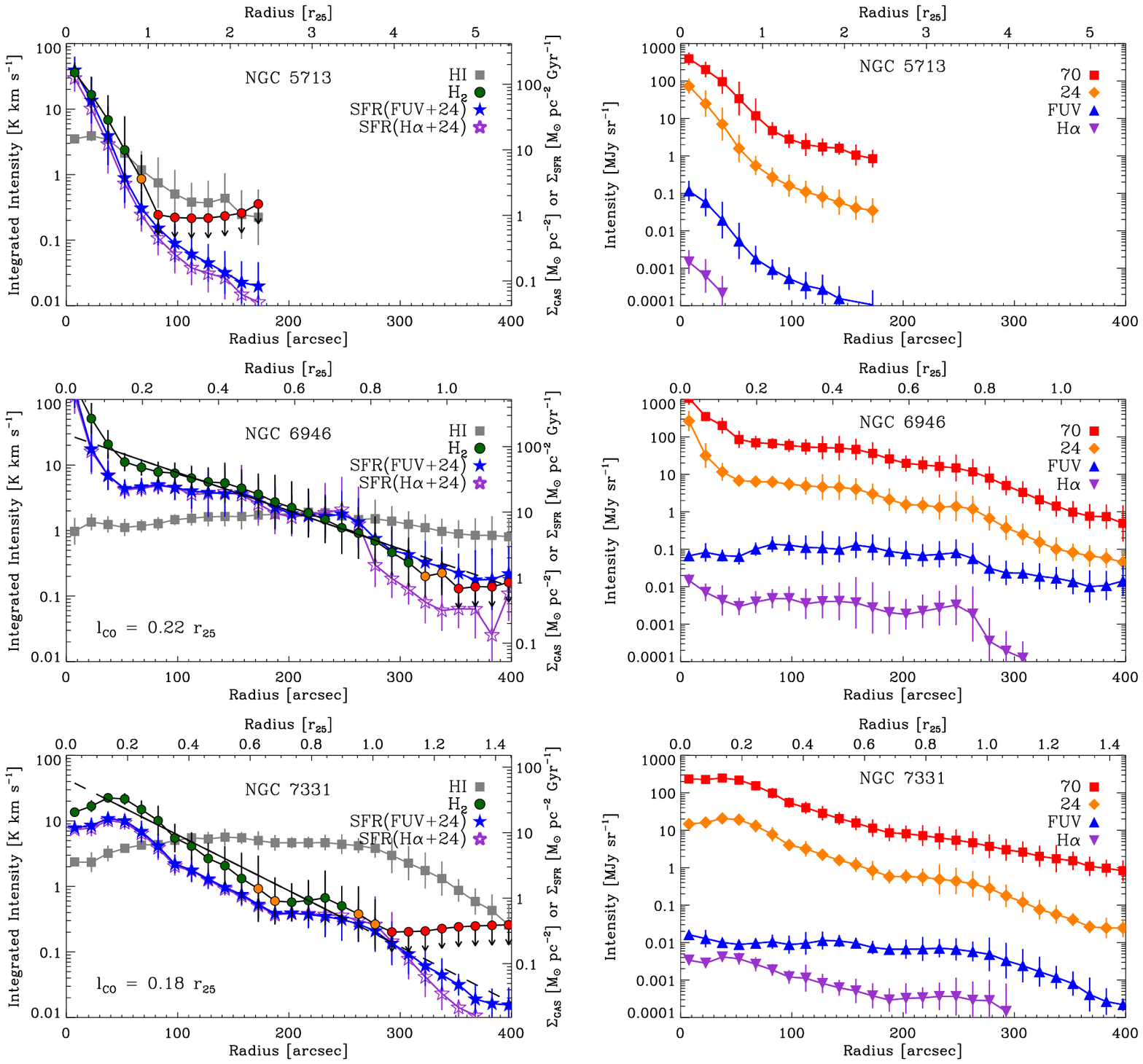}
	\caption{\label{f16g}Atlas of radial profiles, see Figure~\ref{f4} for details.}
\end{figure*}

\end{appendix}

\end{document}